\documentclass{aa}  
\usepackage{color}
\usepackage{natbib,twoopt}
\usepackage{subcaption}
\usepackage{amsmath,amssymb}
\usepackage[breaklinks=true]{hyperref} %% to avoid \citeads line fills
\bibpunct{(}{)}{;}{a}{}{,} %% natbib format for A&A and ApJ
\makeatletter
\newcommandtwoopt{\citeads}[3][][]{\href{http://adsabs.harvard.edu/abs/#3}%
{\def\hyper@linkstart##1##2{}%
\let\hyper@linkend\@empty\citealp[#1][#2]{#3}}}
\newcommandtwoopt{\citepads}[3][][]{\href{http://adsabs.harvard.edu/abs/#3}%
{\def\hyper@linkstart##1##2{}%
\let\hyper@linkend\@empty\citep[#1][#2]{#3}}}
\newcommandtwoopt{\citetads}[3][][]{\href{http://adsabs.harvard.edu/abs/#3}%
{\def\hyper@linkstart##1##2{}%
\let\hyper@linkend\@empty\citet[#1][#2]{#3}}}
\newcommandtwoopt{\citeyearads}[3][][]%
{\href{http://adsabs.harvard.edu/abs/#3}
{\def\hyper@linkstart##1##2{}%
\let\hyper@linkend\@empty\citeyear[#1][#2]{#3}}}

\def\Msun{\ensuremath{\rm \,M_\odot\,}}

\makeatother

\def\mean#1{\left< #1 \right>}

\usepackage{txfonts}
\begin{document}

\title{Diagnostics of the unstable envelopes of \\Wolf-Rayet stars}

   \author{L. Grassitelli
          \inst{1}\fnmsep\thanks{e-mail: luca@astro.uni-bonn.de}
          ,
         A.-N. Chen\'{e}\inst{2}
         ,
         D. Sanyal\inst{1}
         ,
         N. Langer\inst{1}
         ,
         N. St.Louis\inst{3}
         ,      
                  J.M. Bestenlehner\inst{4,1},         
         and
         L. Fossati\inst{5,1}        
}

   \institute{Argelander-Institut f\"ur Astronomie, Universit\"at Bonn, Auf
              dem H\"ugel 71, 53121 Bonn, Germany
         \and
                         Gemini Observatory, Northern Operations Center, 670 North A’ohoku Place, Hilo, HI 96720, USA                 
         \and
         D\'{e}partement de Physique, Pavillon Roger Gaudry, Universit\'{e} Montr\'{e}al, CP 6128, Succ. Centre-Ville, Montréal, H3C 3J7 Quebec, Canada
         \and
              Max-Planck-Institute for Astronomy, 
              Königstuhl 17, 69117 Heidelberg, Germany 
         \and
              Space Research Institute, Austrian Academy of Sciences, Schmiedlstrasse 6, A-8042 Graz, Austria     
         }

%  \date{Received February 29, 2012; accepted --}
   \date{Received //, 2015}

% \abstract{}{}{}{}{} 
% 5 {} token are mandatory
 \abstract
      {The envelopes of stars near the Eddington limit are prone to various instabilities. 
      %We investigate the partly convective envelopes of massive helium stars, where the instability to convection is due to the iron opacity peak and where the envelope in some cases is unstable against pulsations.
A high Eddington factor in connection with the iron opacity peak leads to convective instability, and a corresponding envelope inflation may induce pulsational instability. Here, we investigate the occurrence and consequences of both instabilities in models of Wolf-Rayet stars.       
       } 
   {We determine the convective velocities in the sub-surface convective zones to estimate the amplitude of the turbulent velocity at the base of the wind that potentially leads to the formation of small-scale wind structures, as observed in several Wolf-Rayet stars. We also investigate the effect of stellar wind mass loss on the pulsations of our stellar models.}
   {We approximated solar metallicity Wolf-Rayet stars in the range $\rm 2-17\,M_\odot$ by models of mass-losing helium stars, computed with the Bonn stellar evolution code. We characterized the properties of convection in the envelope of these stars adopting the standard mixing length theory. }
   {Our results show  the occurrence of sub-surface convective regions in all studied models. Small ($\rm \approx 1\, km/s$) surface velocity amplitudes are predicted for models with masses below $\rm \approx 10 \, {M}_\odot$. For models with $\rm{M}\gtrsim 10\,\rm{M}_\odot$, the surface velocity amplitudes are of the order of $\rm 10\, km/s$. Moreover we find the occurrence of pulsations for stars in the mass range 9-14M$\rm _\odot$,  while mass loss appears to stabilize the more massive Wolf-Rayet stars. We confront our results with observationally derived line variabilities of 17 WN stars, of which we analysed eight here for the first time. The data suggest variability to occur for stars above $10 \Msun$, which is increasing linearly with mass above this value, in agreement with our results. We further find our models in the mass range 9-14\Msun to be unstable to radial pulsations, and predict local magnetic fields of the order of hundreds of Gauss in Wolf-Rayet stars more massive than $\approx 10\Msun$.}
  % conclusions heading (optional7)
   {Our study relates the surface velocity fluctuations induced by sub-surface convection to the formation of clumping in the inner part of the wind. From this mechanism, we expect a stronger variability in more massive Wolf-Rayet stars, and a weaker variability in corresponding low metallicity Wolf-Rayet stars.}

      \authorrunning{Grassitelli et al.
}
         
   \maketitle

%       \titlerunning
%
%________________________________________________________________

\section{Introduction}

Wolf-Rayet (WR) stars of spectral class WNE (nitrogen rich) are very hot, highly luminous stars that are thought to be the bare cores of evolved massive stars. Owing to strong stellar winds, these stars have lost almost all their hydrogen-rich envelope allowing us to model them  as H-free helium stars \citep{1986Chiosi,1989Langer}.\\ \indent 
Such stars are expected to develop a convective core, due to the high central energy production, and a radiative envelope \citep{1990Kippenhahn}. %This is true until we approach the surface. 
In addition, the temperature in the outer layers decreases so much that another convective region arises because of a bump in opacity at $\rm{log}(T)\cong5.3$, which is associated with iron and iron-group elements  \citep[FeCZ;][]{1994Langer,1996Iglesias,2012Grafener,2012Langer}. As the opacity increases, the increased radiative acceleration brings the layers close to their local Eddington luminosity and, to avoid exceeding this luminosity, the star is forced to expand forming a low density inflated convective envelope \citep{2006Petrovic,2012Grafener,2015Sanyal}. This convective region is inefficient in transporting energy, but may lead to observable effects, such as turbulence in the surface layers, and may directly influence the wind at its base
\citep{1986Owocki,1996Heger,2005Glatzel,2008Moffat,2009Cantiello,2012Langer,2013Grafener,2015GrassitelliA}.

The mass-loss rate in massive stars is a crucial ingredient in determining correct stellar parameters, the evolutionary scenario, and consequently the fate of these stars \citep{2012Langer}.
In the WR phase the mass-loss rate and wind density are sufficiently high for the stellar winds to be optically thick, obscuring the hydrostatic layers from direct observations.

The spectra of WR stars of  type WNE are dominated by broad emission lines of helium and, in part, nitrogen. Spectroscopic analysis of line profiles reveal the presence of systematic variability in the form of emission sub-peaks migrating from the line centre to the wings \citep{1988Moffat,1999Lepine}. Phenomenological investigations of these discrete wind emission elements involve the presence of a high number of randomly distributed stochastic density inhomogeneities, i.e. clumps in the wind of WR stars. The non-periodical nature of these radially propagating clumps and the migration of the sub-peaks have partially ruled out non-radial pulsation as the origin of these inhomogeneities \citep{1996Cranmer,1999Lepine}.

These structures in the outflow dynamically arise from instabilities that may be triggered at the base of the wind \citep{2015Owocki}. Given the stochastic nature of this phenomenon, we investigate the conditions in which clumping can be triggered by sub-surface convection and can eventually be enhanced during the radial propagation in the wind by radiative instabilities \citep{1988Owocki,2015Owocki}. In fact, observational evidence tends to show that clumps are formed at the base of the wind and do not develop when the wind is already significantly supersonic \citep{2005Davies,2014Vink,2015Owocki}. The phenomenon called clumping affects the parameters derived from model atmosphere analysis introducing further uncertainties in the observational analysis \citep{1994Moffat,2008Hillier}.
Therefore an understanding of the origin of the observed variability and instabilities in the outer layers of these stars is fundamental for correctly estimating mass-loss rates and for understanding the physics of radiation pressure dominated winds \citep{1998Hamann}.

%The effects of the presence of the Fe-bump and pulsations on the outer layers of very massive stars is not well understood. In addition, the majority of the stellar atmosphere codes created for the analysis of cool stars do not take convection into account.

As a result of the buoyancy force the convective elements move upwards and downwards and, in some cases, they may reach velocities of the order of the local sound speed. The interaction with the material in the upper radiative layer may consequently generate gravity waves that propagate to the surface, potentially leading to observable phenomena such as turbulence or small-scale velocity fields, as in the case of micro- and macroturbulence \citep{1990Goldreich,2009Cantiello,2015GrassitelliB,2015GrassitelliA}. The appearance of a velocity field at the surface is connected with the outermost region of the sub-surface convective zone and the physical conditions in the low density envelope. However, observing these perturbations at the stellar surface is hampered by the optically thick wind and only indirect evidence might be inferred \citep{1989Langer,2013Grafener}. 

Another type of spectral variability is present in the wind of some WR stars. This variability is related to the presence of the corotating interaction regions \citep[CIRs;][]{1984Mullan,1986Mullan,1996Cranmer,2002Dessart,2009StLouis,2011CheneStLouis}, connected with the rotation of WR stars of the order of 50 km/s \citep{2003Meynet,2006Yoon,2012Grafener2}. This large-scale variability in the wind of hot massive stars has a periodic behaviour, although this behaviour is epoch dependent, and is thought to be a signature for spiral-like structures appearing in the wind and carried around by rotation \citep{1996Cranmer,2002Dessart,2009StLouis}. The triggering mechanism at the base of the wind is probably not related to stochastic convective motions given the periodical stream-like nature of CIRs. Instead, studies have suggested that CIRs are analogous to the spiral structures in the solar corona and solar wind \citep{1976Hundhausen,1999Gosling,2009StLouis}, which are linked with magnetic fields and most likely also connected with shocks observed in the wind of hot luminous stars \citep{1997Berghofer,2006Marchenko,2008Lepine}. In this paper we neglect to include both CIRs and large-scale variability in our direct analysis, but rather try to enlighten some aspects related to the observed small-scale variability in WR stars.

Furthermore, theoretical works \citep[e.g.][]{1993Glatzel,1994Glatzel,1998Saio} predict the appearance of periodic variabilities arising from pulsations in the envelopes of stars with high luminosity-to-mass ratios. However, up to now, this periodical variability has not been clearly observed and identified, challenging both observations and models. In the context of WR stars, both observation and pulsational analysis need to take into account the effects of mass loss via stellar wind in the outer layers, a key feature in this class of objects. Therefore we also investigate the instability to pulsations of our helium star models in a hydrodynamically consistent way, including the effects of mass loss for the first time.  

We present a set of helium star models computed with a state-of-the-art hydrodynamical stellar evolution code. We present the methods used to compute stellar models and to study the interaction between convective and radiative layers in Sect.2, and the results of our analysis  are shown in Sect.3. In Sect.4 we present the results obtained from the pulsating models,  in Sect.5 we discuss the possible observational signatures comparing our predictions to previously and newly analysed observations, and in Sect.6 we give our conclusions.

%%%%%%%%%%%%%%%%%%%%%%%%%%%%%%%%%%%%%%%%%%%%%%%%%%%%%%%%%%%%%%%%%%%%%%%%%%%%%%%%%%%%%%%%%%%%%5
\section{Method}

\begin{figure}
\resizebox{\hsize}{!}{\includegraphics{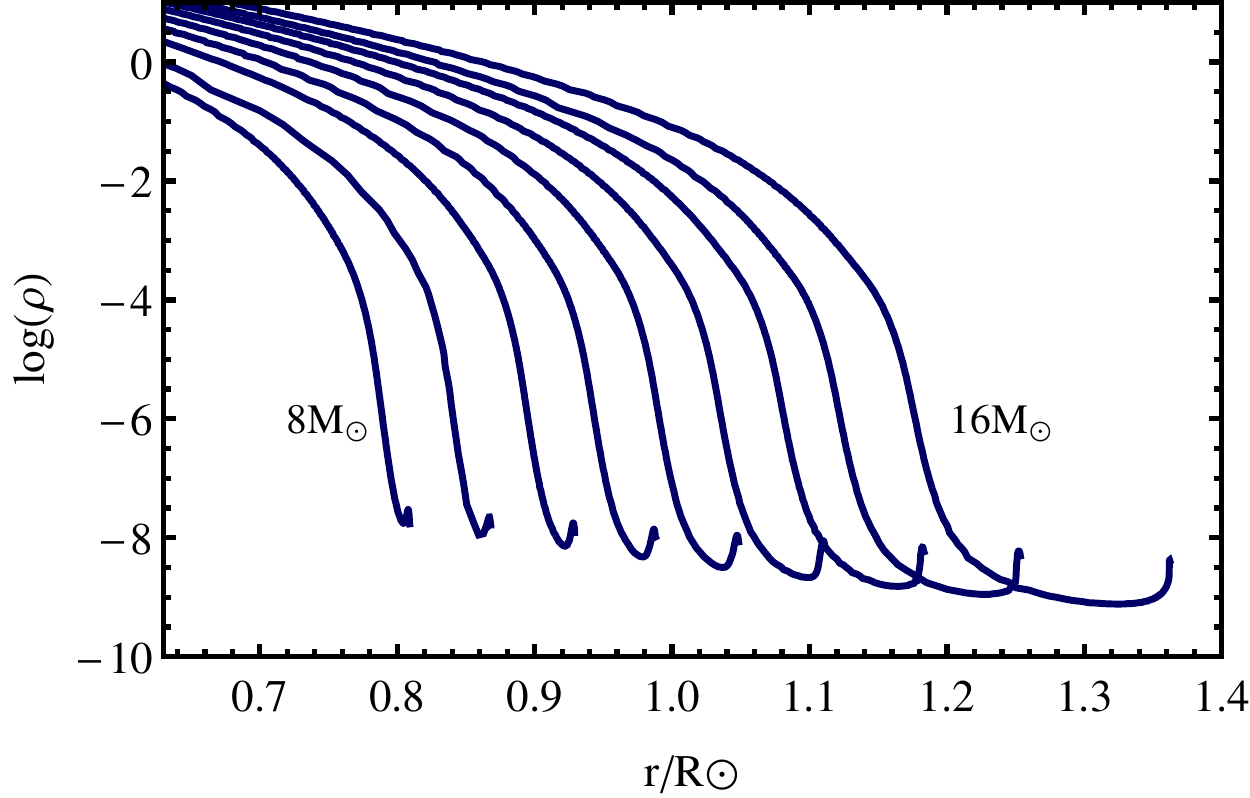}}\caption{ Density profiles (in $g/cm^3$) in the outer regions of our helium stars models with stellar masses ranging from $\rm8\,M_\odot$ to $\rm16\,M_\odot$. The extent of the inflated envelope increases while the density decreases with increasing mass. }
\label{fig:logro}
\end{figure}

\begin{figure}
\resizebox{\hsize}{!}{\includegraphics{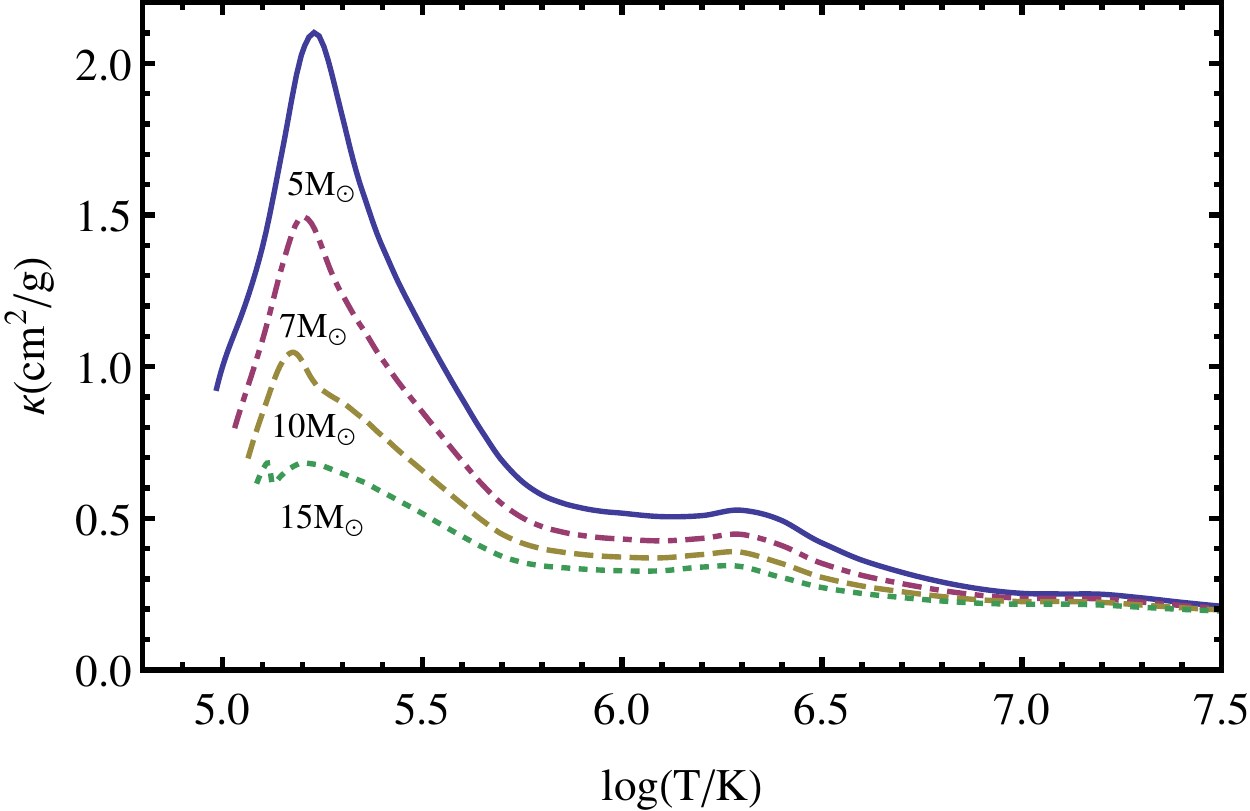}}  \caption{Opacity $\kappa$ as a function of temperature for helium stars of different masses using OPAL opacity tables. The opacity bump from the iron group elements is clearly visible at $\rm log(T/K)\cong 5.3$.}
\label{fig:kappa}

\end{figure}

We computed a set of chemically homogeneous helium star models with masses ranging from $\rm 2M_{\odot}$ to $\rm 17M_{\odot}$ with a one-dimensional (1-D) hydrodynamical stellar evolution code described in \citet{2000Heger}, \citet{2006Petrovic}, \citet{2011Brott}, and references therein. The approach to use chemically homogeneous models is justified since the properties of H-free WN stars are insensitive to the internal chemical structure or evolutionary history \citep{1989Langer}.
The code adopts the standard non-adiabatic mixing length theory (MLT) to model the convective regions and physical quantities associated with the convective layers such as the convective velocities
% or the advection of radiative energy 
\citep{1958Vitense,1990Kippenhahn,2000Heger}. The mixing length parameter $\alpha$ is assumed to be equal to 1.5 as in \citet{1997Abbett}. The models are computed with outer boundary conditions derived from the assumption of a plane parallel grey atmosphere. A consequence of this assumption is that the feedback from an optically thick wind on the outer hydrostatic layers is neglected.
%, which leads to uncertainties which may, to some extent, limit the validity of this approach.
The models are computed with OPAL opacity tables from \citet{1996Iglesias} and with a metallicity of $ \rm Z=0.02$. We adopt the mass-loss rate prescription from \citet{2000Nugis} and neglect the effects of rotation and magnetic fields. 

In order to determine whether a layer is unstable to convection, we adopt the Schwarzschild criterion, which can be expressed as
\begin{equation}
\label{schcrit}
\Gamma = \frac{\kappa L}{4 \pi c G M} \geq (1-\beta)\frac{32 - 24 \beta}{32 - 24 \beta - 3 \beta^2}
,\end{equation}
where $\beta$ is the ratio of gas pressure to total pressure, $M$ is the stellar mass, $L$ the luminosity of the star, $G$ the gravitational constant, $\kappa$ the Rosseland mean opacity, and $\Gamma$ the local Eddington factor \citep{1973Joss,1997Langer}. 

Below, we compare the convective velocities $v_c$ with the isothermal sound speed  to define the Mach number $M_c$.
% In fact, the condition for isentropic perturbations is not fulfilled in such an high opacity-low density region. 
The low density inflated envelope is defined as the layers above the point at which $\beta$ is equal to 0.15 \citep{2015Sanyal}. The comparison with the isothermal sound speed in the envelope can be justified by analysing the timescales in the radiation pressure dominated envelopes. The thermal timescale is\begin{equation}
\tau_{th} = \frac{G M \Delta M_{env}}{R L}
,\end{equation}where $ \Delta M_{env}$ is the mass of the inflated envelope and $R$ the stellar radius. This timescale is of the order of 100\,s in our set of models (see Table \ref{tab2}), whereas the dynamical timescale $\rm \tau_{dyn}$, which is written as $ \tau_{dyn}=\sqrt{R^3/G M}$, is one order of magnitude larger. This suggests that a perturbation would rapidly thermalize, therefore the isothermal sound speed is thought to be a better approximation for the real sound speed in these kinds of models.  

%The assumption of an isothermal sound speed in the envelope supports the idea of both gravity and acoustic waves propagating upward from the convective region, as described in \citet{1990Goldreich} plus acoustic pulsation modes of not thermal origin \citep{1994Glatzel,2003Blaes}. 
%Thus, solutions of the wave equation in isothermal and stratified outer layers are proportional to $\sim$ Exp$(-k_{\pm} r)$, where $r$ is the radial coordinate and the real part of $k_{\pm}$ is equal to $(2H_p)^{-1}$.\\ 
%To estimate the correlation length  we assume that it corresponds to the pressure scale height at the surface as 
\citet{2009Cantiello} presented a study of the convective regions in the outer envelopes of massive main-sequence stars and of the induced velocity field at the surface via propagation of uncorrelated gravity waves. Following \citet{2009Cantiello}, we introduce $\mean{v_c}$ as the convective velocity averaged over one mixing length starting from the upper boundary of the sub-surface convective zone at $\rm r = R_c $, defined as in \citet{2009Cantiello},
\begin{equation}\label{convlasthp}
\mean{v_c}:= \frac{1}{\alpha H_P} \int^{R_c}_{R_c-\alpha H_P} v_c(r)\,dr \quad,
\end{equation}
where the mixing length $\alpha H_P$ is obtained by the product of the pressure scale height $H_P$  (with the pressure given by the sum of gas and radiation pressure) and $\alpha$. Gravity waves propagate typically with a frequency $\omega \sim M_c \omega_{ac}$, where $\omega_{ac}$ is the acoustic cut-off frequency, which is of the order of $10^{-2}$Hz in our models, while acoustic waves are expected to have frequencies $\omega > \omega_{ac}$. \citet{1990Goldreich} showed that it is possible to estimate the fraction of convective energy flux going into acoustic and gravity waves from \begin{equation}\label{acugrav}
F_g=M_c\,F_c \quad ,\quad F_a=M_c^{\frac{15}{2}}\,F_c \quad ,
\end{equation}
where $F_c$ is the total convective energy flux and $F_a$ and $F_g$ are the energy fluxes transported by acoustic and gravity waves, respectively. As a result, in the hydrostatic sub-sonic regime (where $M_c <1$) we expect most of the convective energy to go into gravity waves.
%The convective motion of the bulk elements excites acoustic and gravity waves once they interact with the overlying radiative layer, as described in \citet{1990Goldreich,2009Cantiello}. The waves can propagate outward until they reach the surface, carrying an energy that, at the surfuce, has to be of the same order or smaller than the kinetic energy at the upper border of the Fe convective region \citep{2009Cantiello}. \\

In order to investigate the influence of the sub-surface convective zone in helium stars and relate the convective motion to observational phenomena at the surface  without the ability to compute  the energy dissipation by the gravity waves along the path exactly, we introduce an upper limit to the expected velocity amplitude at the surface based on the conservation of energy, namely
\begin{equation}\label{vs}
v_{surf}\leq \mean{v_c} \sqrt{M_c \frac{\rho_c}{\rho_{s}}}  \quad,
\end{equation}where $\rho_s$ and $\rho_c$ are the densities at the surface and at the upper border of the convective zone and where the Mach number is defined as \begin{equation}
M_c=\frac{\mean{v_c}}{c_{T}} \quad ,
\end{equation}
where $c_T$ is the isothermal sound speed at the top of the FeCZ \citep{1990Goldreich,2009Cantiello}.

\section{Sub-surface convection}

\begin{figure*}[!]
\begin{sidecaption}
%\sidecaption
\resizebox{0.7\hsize}{!}{
\includegraphics{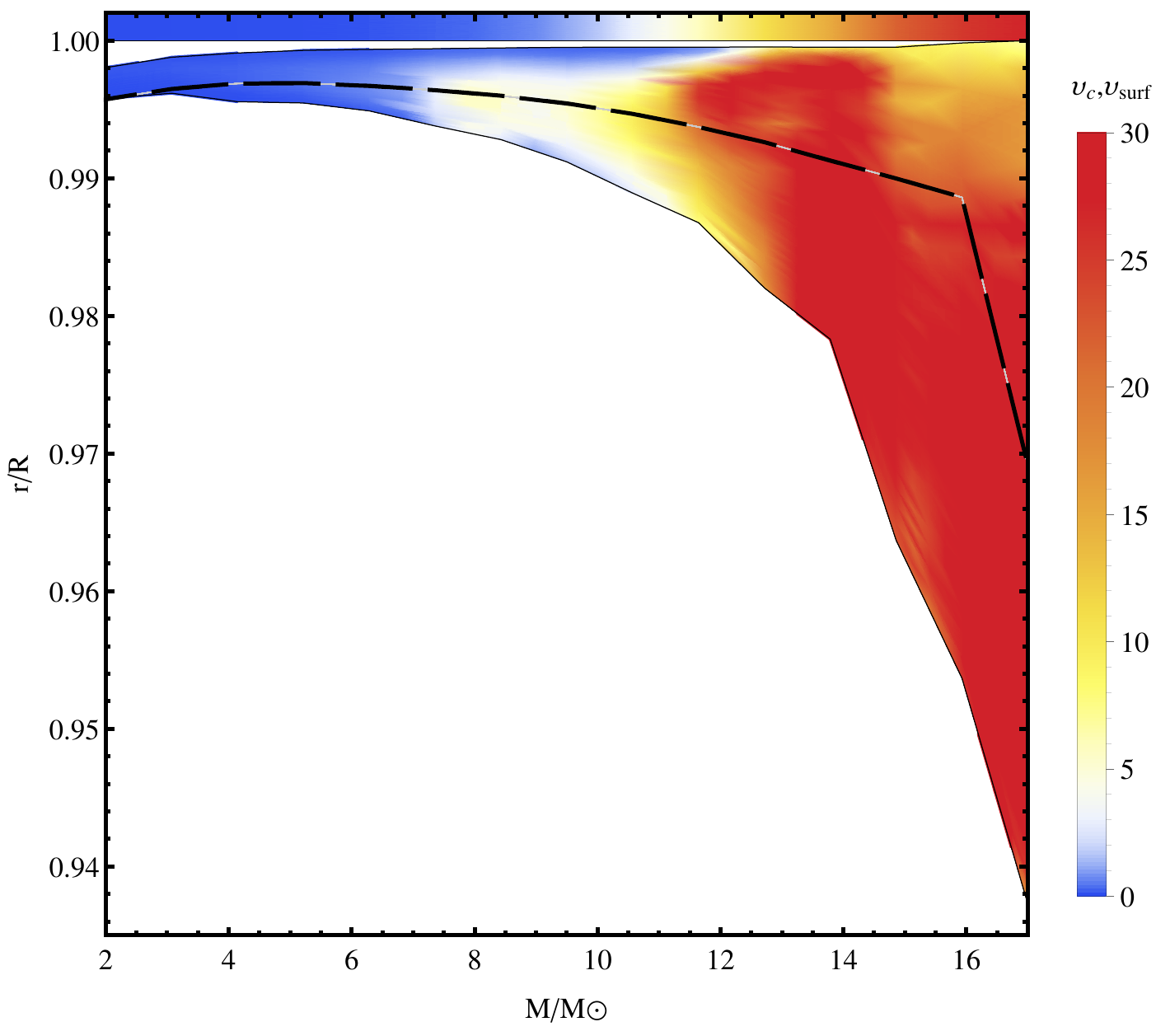} }
\caption{
Normalized radial coordinate as a function of the mass of our helium star models for the outer 6.5\% of the stellar radius. The white region represents radiative layers, while the coloured region denotes convection. The colours indicate the convective velocities within the FeCZ (in km/s); the black dashed line superposed lies one mixing-length away from the upper border of the convective region. The colours in the top bar indicate the expected velocity fluctuations at the surface (\rm see also Fig.\ref{surfconv} for a direct comparison between $\mean{v_c}$ and $v_{surf}$ ).    
%Colour representation of the convective velocities within the convective zone and expected fluctuations at the surface (top-bar). The black dashed line superposed lies one pressure scale height below the upper border of the convective region.
}
\label{densityplot}
\end{sidecaption}
\end{figure*}

\begin{table*}[t]
\centering

\begin{minipage}{\textwidth}   

\caption{Properties of our set of helium zero age main-sequence models with Z=0.02. 
$\rm \Delta M_{env}$ mass of the envelope, $\rm \mean{ v_c}$ averaged convective velocity over the last mixing length of the FeCZ, $\rm v_{surf} $ surface velocity fluctuation, $\rm \Delta R_{conv}$ spatial extent of the convective zone, $\rm M_c $ Mach number for the averaged convective velocity,  $\rm \tau_{conv}$ convective timescale, $\rm B$ magnetic field strength. 
%Mass-loss rate from Nugis $\&$ Lamers (2000). Mass of the inflated envelope defined from the point at which $\beta=0.15$ up to the surface. Blank slots are models who do not reach $\beta=0.15$ in their envelope and we do not define them as inflated.Spatial extent of the iron sub-surface convective zone.time-scale associated with sub-surface convection for the most massive models, see Sect.5.
}

\centering
\begin{tabular}{ l  l  l  l  l  l  l  l  l  l  l r }

%\vspace{0.2cm}
%\vspace{0.2cm}

  \hline                        
\hline  

  $\rm M$ & log $L/L_\odot$ & log $T_{\rm eff}/K$ & $\rm R$ & log $ \dot{M} $  & $\rm \Delta M_{env}$  & $\rm \mean{ v_c}$ & $\rm v_{surf} $ & $\rm \Delta R_{conv}$  & $\rm M_c $ & $\rm \tau_{conv}$ & $\rm B$ \\

    $M_\odot$ & $ $ & $ $ & $R_\odot$ & $M_\odot/yr$ & $10^{-9}M_\odot$ & $km/s$ & $km/s $ & $R_\odot$ & $  $ & $s$ & $G$\\
  
\hline

2 & 3.36 & 4.84 & 0.34 & -6.44  &          & 0.01 & 1.$\times 10^{-3}$  & 7.70$\times 10^{-4}$ & 5.$\times 10^{-4}$ & 51284       & 0.31\\
3 & 4.07 & 4.93 & 0.49 & -6.37  &          & 0.03 & 2.$\times 10^{-3}$  & 1.47$\times 10^{-3}$ & 1.$\times 10^{-3}$  & 35047      & 1.45\\
4 & 4.22 & 4.95 & 0.53 & -6.33  &          & 0.06 & 4.$\times 10^{-3}$  & 1.84$\times 10^{-3}$ & 2.$\times 10^{-3}$  & 20930      & 2.10\\
5 & 4.46 & 4.98 & 0.61 & -6.03  &       & 0.10 & 9.$\times 10^{-3}$  & 2.28$\times 10^{-3}$ & 4.$\times 10^{-3}$ & 14328   & 4.07\\
6 & 4.66 & 5.00 & 0.68 & -5.79  & 0.21  & 0.19 & 0.02  & 2.85$\times 10^{-3}$ & 0.02   & 9670         & 7.49\\
7 & 4.81 & 5.03 & 0.75 & -5.64  & 0.43  & 0.33 & 0.03  & 3.88$\times 10^{-3}$ & 0.01  & 6788  & 12.7\\
8 & 4.94 & 5.04 & 0.81 & -5.50  & 0.63  & 0.60 & 0.11  & 5.08$\times 10^{-3}$ & 0.02    & 4553        & 22.8\\
9 & 5.03 & 5.05 & 0.87 & -5.36  & 1.15  & 1.13 & 0.27  & 6.29$\times 10^{-3}$ & 0.04    & 3003                & 40.7\\
10 & 5.13 & 5.06 & 0.93 & -5.34  & 1.32 & 2.09 & 0.66  & 8.68$\times 10^{-3}$ & 0.07    & 2017                & 70.5\\ 
11 & 5.21 & 5.07 & 0.99 & -5.13  & 1.68 & 3.94 & 1.66  & 1.12$\times 10^{-2}$ & 0.13    & 1343        & 122   \\
12 & 5.29 & 5.07 & 1.05 & -5.04  & 2.23 & 7.01 & 3.89  & 1.42$\times 10^{-2}$ & 0.25    & 949                 & 203   \\
13 & 5.35 & 5.08 & 1.11 & -4.95  & 2.71 & 10.7 & 7.25  & 2.07$\times 10^{-2}$ & 0.37    & 779                 & 284   \\
14 & 5.41 & 5.08 & 1.18 & -4.87  & 3.52 & 15.6 & 12.7  & 2.61$\times 10^{-2}$ & 0.56   & 663          & 387   \\
15 & 5.46 & 5.08 & 1.25 & -4.81  & 4.11 & 18.3 & 16.1  & 4.62$\times 10^{-2}$ & 0.65   & 687          & 415   \\
16 & 5.53 & 5.08 & 1.36 & -4.72  & 5.33 & 22.8 & 21.4  & 6.80$\times 10^{-2}$ & 0.88   & 732          & 532   \\
17 & 5.56 & 5.08 & 1.42 & -4.68  & 5.80 & 26.2 & 25.8  & 8.64$\times 10^{-2}$ & 0.96   & 1619         & 369   \\

  \hline  
  
\end{tabular}

\label{tab1}

\end{minipage}

\end{table*}

Our helium star models with masses ranging from 2 to 17 $\rm M_\odot$ are within the luminosity range $\rm log(L/L_\odot)\approx 4-5.5$ and the temperature range $\rm log(T_{eff}/K)\approx 4.9-5.1$ (Table \ref{tab1}), which is very close to the helium zero age main-sequence computed by \citet{2006Petrovic} and by \citet{2012Grafener}. We encounter numerical difficulties in treating models more massive than 17 $\rm M_\odot$, especially in connection to mass loss by stellar wind and, therefore, we do not investigate them. We find inflated envelopes in the outer layers of models with $ \rm M\geq 6\, M_\odot$, which become more extended with increasing luminosity-to-mass ratios. For nearly all inflated regions we find $\Gamma(r)\approx 1$ \cite[ see Fig.\ref{fig:logro}]{2012Grafener,2015Sanyal}. 

All computed models have a convective region close to the surface extending over a significant fraction of the stellar radius ($\lesssim 10\%$) and that comprises a very small amount of mass, i.e. less than $\rm 10^{-8}\,M_{\odot}$ (see Table \ref{tab1}). This convective region arises because of the opacity bump at  $\rm log(T/K)\cong5.3$ (see Fig.~\ref{fig:kappa}).
An estimate of the relative amount of flux carried by convection \citep{1990Kippenhahn} in our models can be given via
\begin{equation}
\frac{F_c}{F_{tot}}=\frac{4 \,\rho\, c_P T v_c^{\,3}}{\alpha H_P \,g\, F_{tot}} \quad,
\end{equation} 
where $F_c$ and $F_{tot}$ are the convective and total flux, respectively, $c_P$ is the specific heat at constant pressure, $T$ is the temperature, $\rho$ is the density, and $g$ is the gravitational acceleration. The fraction of convective flux decreases when the convective layers are increasingly radiation pressure dominated, i.e. with increasing stellar mass, going from $10^{-6}$ in the low mass to $10^{-10}$ in the more massive helium star models of our sample. The partial ionization zones due to helium and hydrogen recombination are absent as the surface temperatures of our models are $\rm log(T_{eff}/K)\geq 4.9$ (see Fig.\ref{fig:kappa}).

%\approx 10^{-10}\ldots10^{-6}

Our models show that the convective velocities within the sub-surface convective zone are a function of the stellar mass and of the radial coordinate within the convective zone. Figure \ref{densityplot} shows the presence of a convective zone in all the investigated models, whose spatial extent gradually increase for higher masses. 

The helium star models up to $\rm 10 \, M_\odot$ display a convective zone that extends over less than $1\%$ of the stellar radius and with convective elements that reach on average velocities not higher than few km/s. The models with $\rm M\geq 10\, M_\odot$ have convective zones that show convective velocities of the same order as the local sound speed, i.e. velocities up to 30 km/s. Figure \ref{densityplot} shows how convection moves deeper inside in normalized radial coordinate as the envelope becomes larger, and, therefore, it occupies a larger fraction of the stellar radius. The convective region also moves closer to the surface, reaching it in the $\rm 17 \, M_\odot$ model.     

The increase in size of the convective zone is connected to the larger inflated envelope when moving towards higher luminosity-to-mass ratios \cite[ see Fig.\ref{fig:logro}]{2006Petrovic,2015Sanyal}. The envelopes become more and more dominated by radiation pressure dominated for higher mass stellar models, density, and $\beta$ decrease while $\Gamma(r)$ stays approximately equal to unity and the Schwarzschild criterion for convection  (see Eq.\ref{schcrit}) is fulfilled in a larger region. The convective velocities also significantly increase for higher mass models, with the maximum occurring in the lowest density regions (see Fig.\ref{fig:logro}), i.e. where $\beta$ is at its minimum. 
 
\begin{figure}
\resizebox{\hsize}{!}{\includegraphics{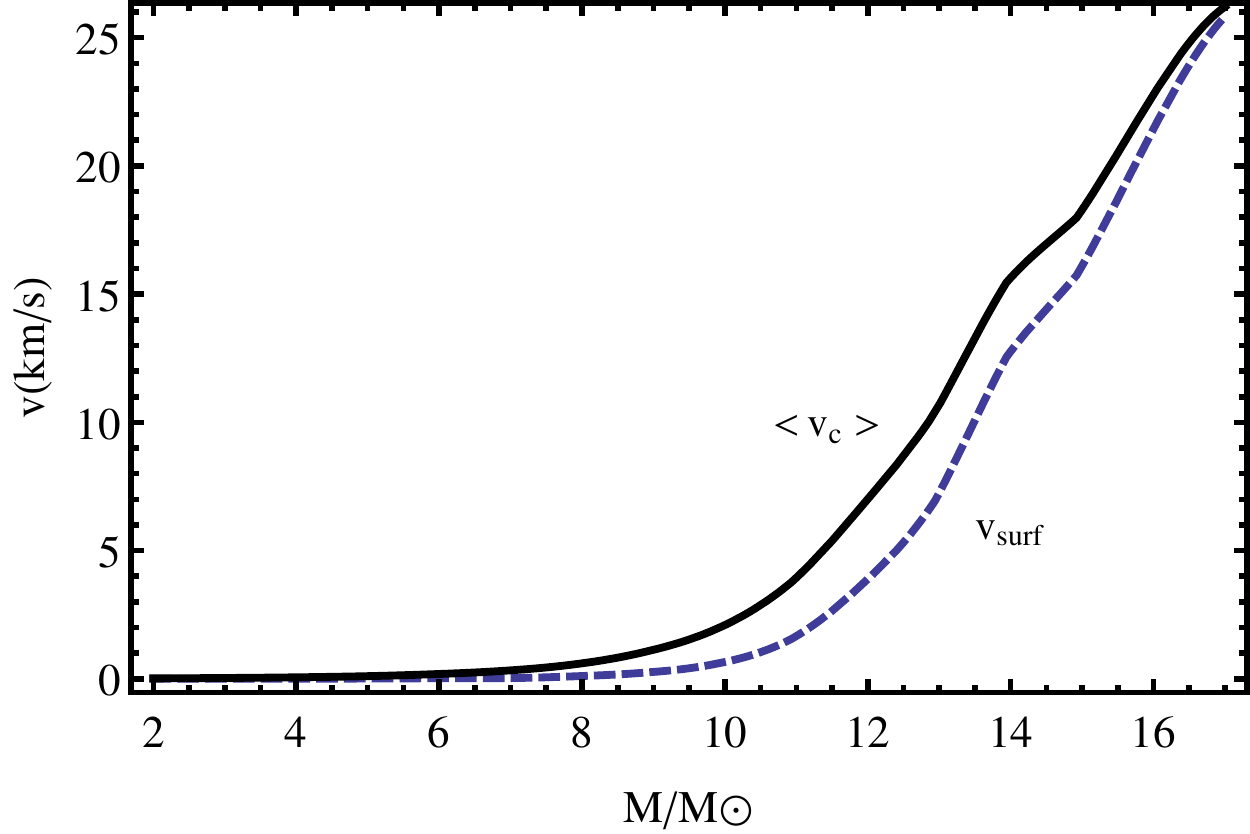}} \caption{Average velocity (black solid line, $\rm \mean{v_c}$) of the convective elements at the top of the convective zone and expected surface velocity fluctuations (blue dashed line, $\rm v_{surf}$) as a function of stellar masses.}
\label{surfconv}
\end{figure}

We compute the average velocity of the convective elements in the upper part of the sub-surface convective zone (Eq.\ref{convlasthp}), assuming implicitly that only the elements in the last mixing length can interact with the upper radiative layer (Fig.\ref{densityplot}).
The models show that these average velocities are smaller than 2 km/s for stars with masses below $\rm 10\,M_{\odot}$. For models with $\rm M\geq 10\,M_{\odot}$, $\mean{v_c}$ becomes larger with values of the order of $20$ km/s in the case of $\rm 15\,M_{\odot}$ (Fig.~\ref{surfconv}). 
%Once defined the Eddington luminosity as:
%\begin{equation}\label{eddington}
%L_{Edd}=4\pi c G \frac{1-\beta}{k} M
%\end{equation} with $c$ the speed of light, $G$ the gravitational constant, $k$ the opacity and $\beta$ the ratio between the gas pressure and the total pressure,
This result can be related again with the fact that when we approach higher stellar masses, the luminosity-to-mass ratios increase. This inflates further the stellar envelope which in turn increases the mixing length and, combined with the higher luminosities,  increases the convective velocities \cite[see Fig.~\ref{fig:logro}, ][]{1990Kippenhahn,2015Sanyal}. This is in agreement with \citet{2009Cantiello} and \citet{2015GrassitelliA}, who find that convective velocities in the FeCZ are higher for higher luminosities, while here, conversely from main-sequence stellar models, the convective velocities are in general higher for higher effective temperatures.

Figure \ref{surfconv} shows the average convective velocities at the top of the iron convective zone and the expected velocity perturbation at the surface. We can see how the velocities are attenuated in the radiative zones, i.e. the expected velocity fields at the surface are smaller than $\mean{v_c}$. This is due to the factor $\sqrt{M_c\, \rho_c/\rho_{s}} \textless 1$ in Eq.\ref{vs}. However the expected $v_{surf}$ roughly follow the increase of the convective velocities in the case of the most massive models. Moreover in the models computed with a stellar mass larger than $\rm 12\, M_\odot$, the relative attenuation becomes lower and the difference between the average convective velocities and surface velocity fields is reduced. In fact the convective layers approach the surface and the radiative layers are less extended. Both the Mach number and  ratio $\rho_c/\rho_s$ approach unity in the models more massive than $\rm 15 M_\odot$ with the convective zone reaching the surface in the $\rm 17 M_\odot$ model. In this configuration one can forsake the assumption made in Eq.\ref{vs} concerning the propagation of the waves, given the absence of a radiative layer separating the convective region from the surface.   
%by adopting a very simplistic approach to take into account the propagation of 
%the waves travelling in the radiative zone that the waves are partially damped when they reach the surface. 
%On a relative scale the attenuation in the helium star models is significantly smaller in percent than in the case of the main sequence stars investigated by \citet{2009Cantiello}. This is due to the much smaller radiative layers separating the convective zone from the surface and to the Mach numbers that are in general higher in this type of stars compared to the main sequence stars. This also makes the velocity field at the surface in the helium star models much more intense respect to that of the main sequence star models in \citet{2009Cantiello}. Moreover, having adopted the isothermal sound speed and not the adiabatic sound speed to define the Mach number increases  significantly the Mach number. Accordingly, damping is less efficient, in contrast to that case of main sequence stars where the adiabatic sound speed have been adopted \citep{2009Cantiello}. 
%In the models with a stellar mass smaller than $\rm 13 \,M_\odot$ the waves are significantly damped when they reach the surface. For these models the Mach number is $ M_c \ll 1$ (Eq.\ref{vs}).   

We thus infer from Fig.\ref{densityplot} and Fig.\ref{surfconv} that low-mass helium stars are not expected to show strong velocity fields originating in the iron convective zone at the base of the wind. A steeper trend of increasing surface velocity fluctuations is expected for helium stars with higher luminosity-to-mass ratios due to the higher convective velocities in the region of the iron bump and the lower distance from the surface. An extrapolation of these results to higher masses ($\rm \geq 17 M_\odot$) needs to be performed with caution. This is because the computed convective velocities are expected to become supersonic and the standard MLT may not apply anymore \citep{1990Goldreich,1991Canuto}. 

The number of clumps expected to be triggered by these perturbations may be roughly estimated, assuming a transversal correlation length at the surface of the order of the local pressure scale height \citep{2009Cantiello}.  This is usually of the order of $10^9$ cm in our massive models, a scale that is very similar to the lateral coherence scale of few degrees that is able to reproduce observations by \citet{2002DessartOwocki} while investigating the line profile variability of small-scale structures in the wind of WR stars. With this approach, we estimate a number of clumps $N_{Clumps}$,
\begin{equation}\label{nclumps}
N_{Clumps}=\frac{4 \pi H_P^2}{4 \pi R^2}\approx 10^3-10^4
,\end{equation}
with a decrease in number for the more massive models.  
   
In summary, convective velocities of the order of the local sound speed have been found in the case of the highest mass models considered here. A steep increase in $\mean{v_c}$ has been found starting from $\rm \approx 10\,M_\odot$. Similarly, the expected velocity field at the surface sharply increases for $\rm M > 10\,M_\odot$, approaching the local sound speed for the most massive models.       
    
%By the use of Eq.\ref{vs}4
%Fig.\ref{densityplot} shows that the convective velocities start to become significant for masses $M>8M_{\odot}$, reaching $10s$ of km/s around $15M_{\odot}$. 
%Furthermore, considering the velocities at the surface, Fig.\ref{surfconv}, the results show that their values lie below the values of $\langle v_c \rangle$, i.e. the perturbations are partially damped travelling through the radiative layer, Eq.\ref{vs}. This in turn makes the velocity field amplitude at the surface to be shifted towards higher masses respect to the averaged convective velocities and to become significant for $M>10M_{\odot}$. \\
%An estimate of the turbulent pressure at the surface due to the induced velocity field in the case of the 15M$_\odot$ model can be computed as:
%\begin{equation}\label{pres}
%\frac{P_{tur}}{P_{tot}}=\frac{\frac{1}{3}\rho_s v_s^2}{P_{tot}}\approx 0.01
%\end{equation}
%with P$_{tot}$ the total pressure at the surface and P$_{tur}$ turbulent pressure \citep{2009Maeder}.

%\newpage

\section{Pulsations}

\begin{table*}

\begin{minipage}{\textwidth}
\centering
 \caption{Physical parameters of the models that are unstable to pulsations.
   Luminosity difference, $\rm \Delta L$, between the maximum and minimum radii, $\rm \Delta m $ bolometric magnitude difference between maximum and minimum radii, $\rm P$ pulsation period, $\rm \Delta R$ difference between maximum and minimum radii, $\rm v^{max}_{osc}$ maximum velocity achieved during a pulsation cycle, $\rm \tau_{dyn}$ dynamical timescales, $\rm \tau_{th}$ thermal timescales,  and $\rm \omega_{dyn}$ frequency of the pulsations normalized by the dynamical timescale as defined in \citet{1993Glatzel}.  
 }

\begin{tabular}{ l  l  l  l  l  l  l  l  r }
  \hline
  \hline                        
  $\rm M$ & $\rm \Delta L$  & $\rm \Delta m $ & $\rm P$ & $\rm \Delta R$ & $\rm v^{max}_{osc}$ & $\rm \tau_{dyn}$ & $\rm \tau_{th}$ & $\rm \omega_{dyn}$  \\
    $M_\odot$ & $L_\odot$ & $10^{-7}mag$ & $s$ & $R_\odot$ & $km/s$ & $s$ & $s$ & $ $ \\ 
\hline 
  9 &  0.00038 & 0.03 & 220 & 4$\times10^{-4}$ & 7.74   & 429 & 109 & 7.08  \\
  10 & 0.01694 & 1.33 & 270 & 8$\times10^{-3}$ & 74.0  & 451 & 101 & 6.06  \\
  11 & 0.04082 & 2.67 & 330 & 0.01 & 101 & 471 & 112 & 5.18  \\
  12 & 0.08167 & 4.52 & 400 & 0.02 & 130 & 493 & 130 & 4.47  \\
  13 & 0.05357 & 2.55 & 459 & 0.03 & 153 & 517 & 138 & 4.09  \\
  14 & 0.02262 & 0.94 & 550 & 0.01 & 59.3  & 545 & 159 & 3.60  \\
  \hline  
\end{tabular}

\label{tab2}

\end{minipage} 

\end{table*}

\begin{figure}\resizebox{\hsize}{!}{\includegraphics{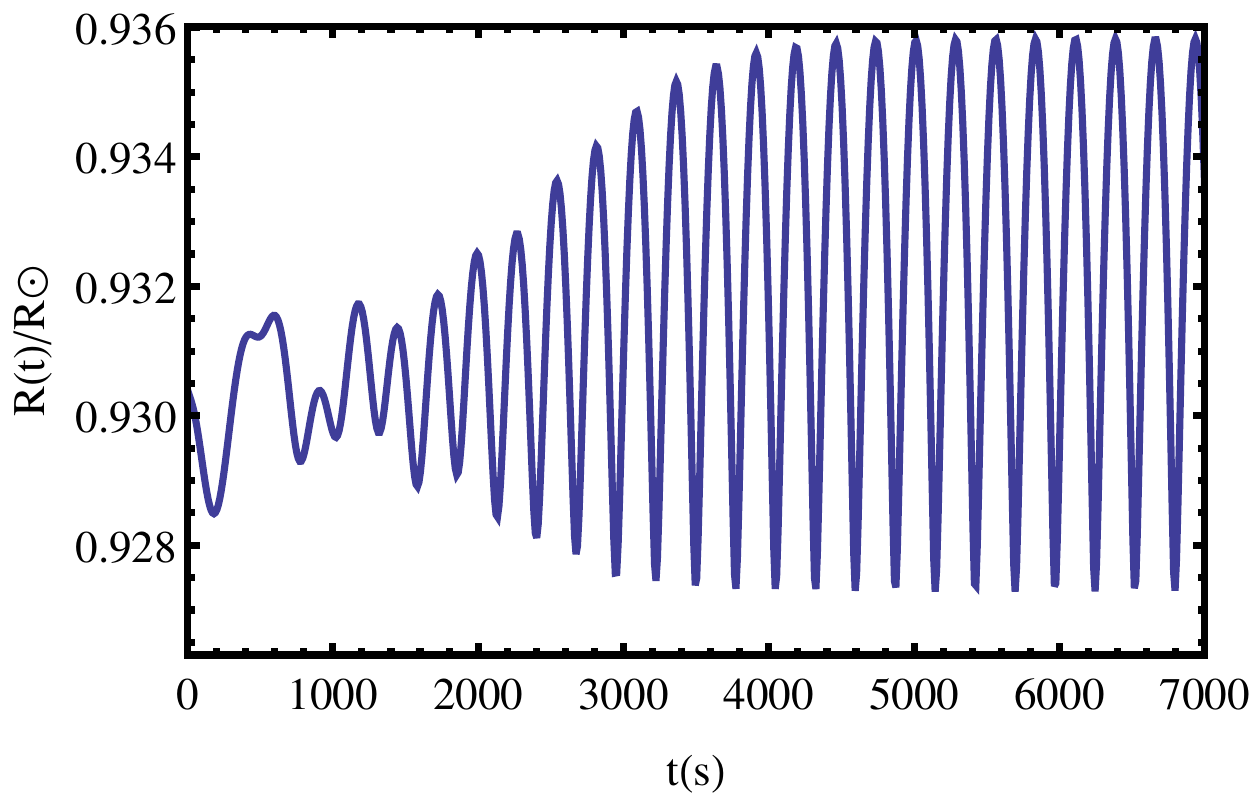}}\caption{Evolution of the radius as a function of time for a 10$\rm M_{\odot}$ radially pulsating helium star model. The pulsation amplitude grows rapidly before saturating.}
\label{fig:pulgro}
\end{figure}

\begin{figure}
\resizebox{\hsize}{!}{\includegraphics{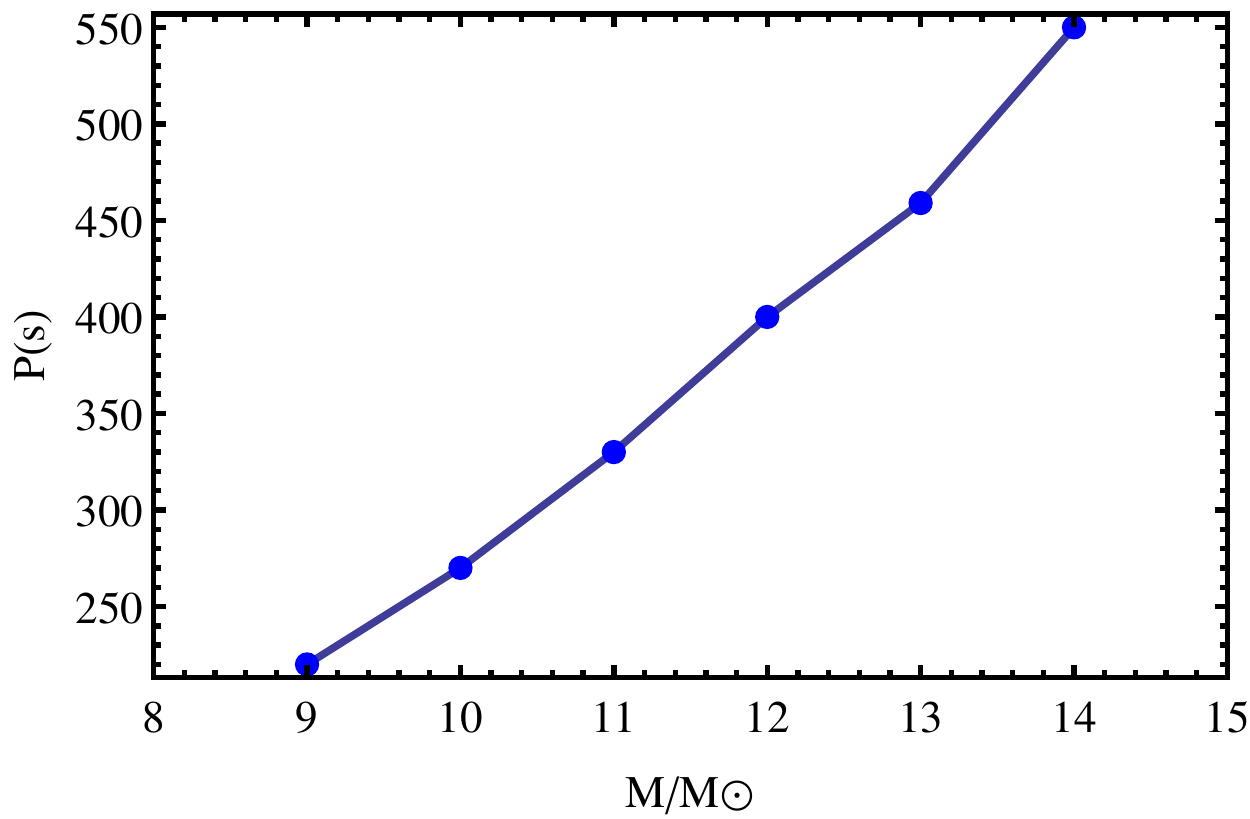}}\caption{Period of a saturated pulsational cycle for different pulsating model masses. }
\label{fig:period}
\end{figure}

\begin{figure}
\resizebox{\hsize}{!}{\includegraphics{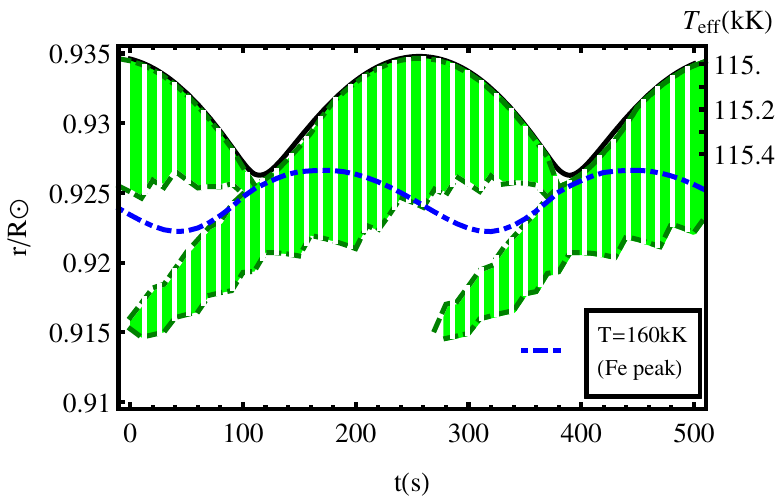}}\caption{Radial coordinate and surface temperature as a function of time for a 10$\rm\, M_\odot$ pulsating stellar model. The black thick line is the surface of the stellar model and the green striped region shows the layers unstable to convection during a pulsation cycle. The blue dot-dashed line tracks the location as a function of time of the temperature at which the iron opacity peak is maximum in a static configuration, i.e. $\rm log(T/K) \approx 5.3$ (see also Fig.\ref{fig:kappa}).  }
\label{fig:convpuls}
\end{figure}

\begin{figure}
\resizebox{\hsize}{!}{\includegraphics{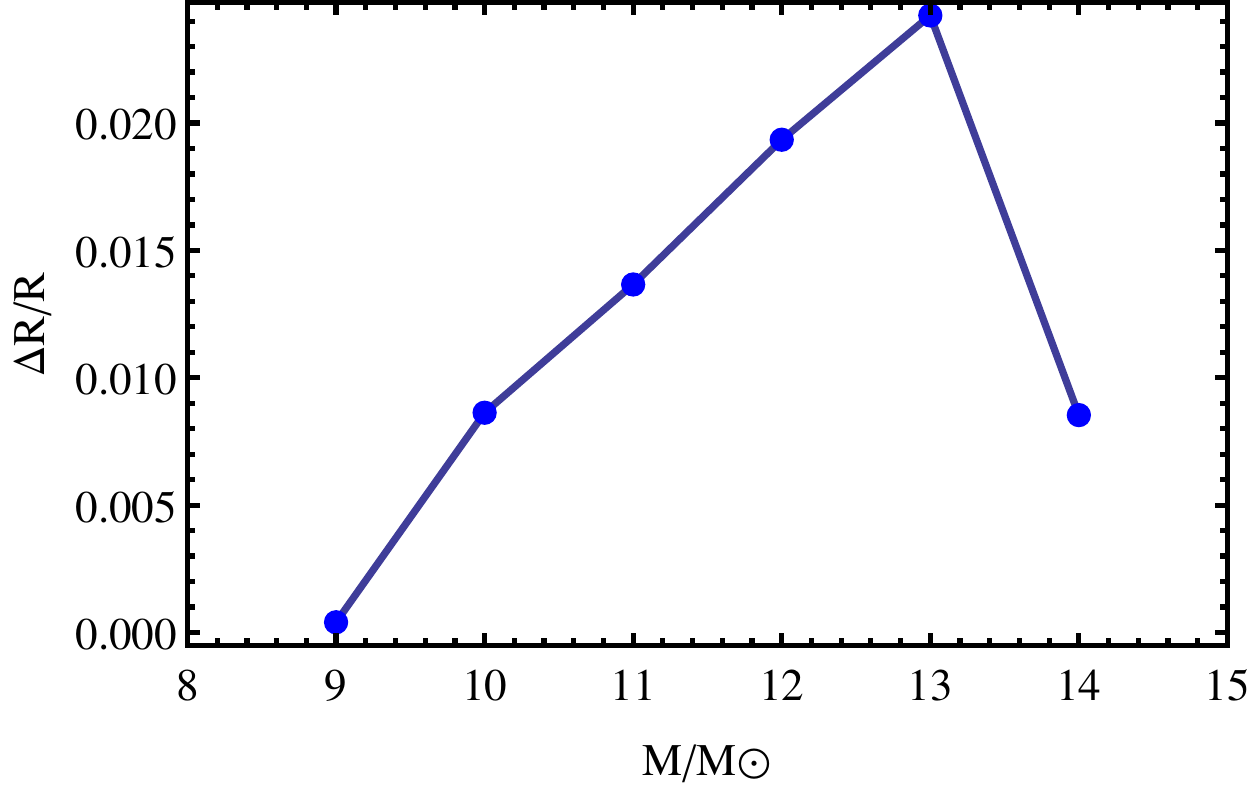}}\caption{Saturated amplitude of the pulsations appearing in the helium star models. 
%The increase in the low mass part of the plot is connected with the growth of the extent of the inflated low density envelopes, while 
The decreasing amplitude above 13$\rm M_\odot$ is believed to be an effect of the increased mass-loss rate. }
\label{fig:ampl}
\end{figure} 

\begin{figure}
\resizebox{\hsize}{!}{\includegraphics{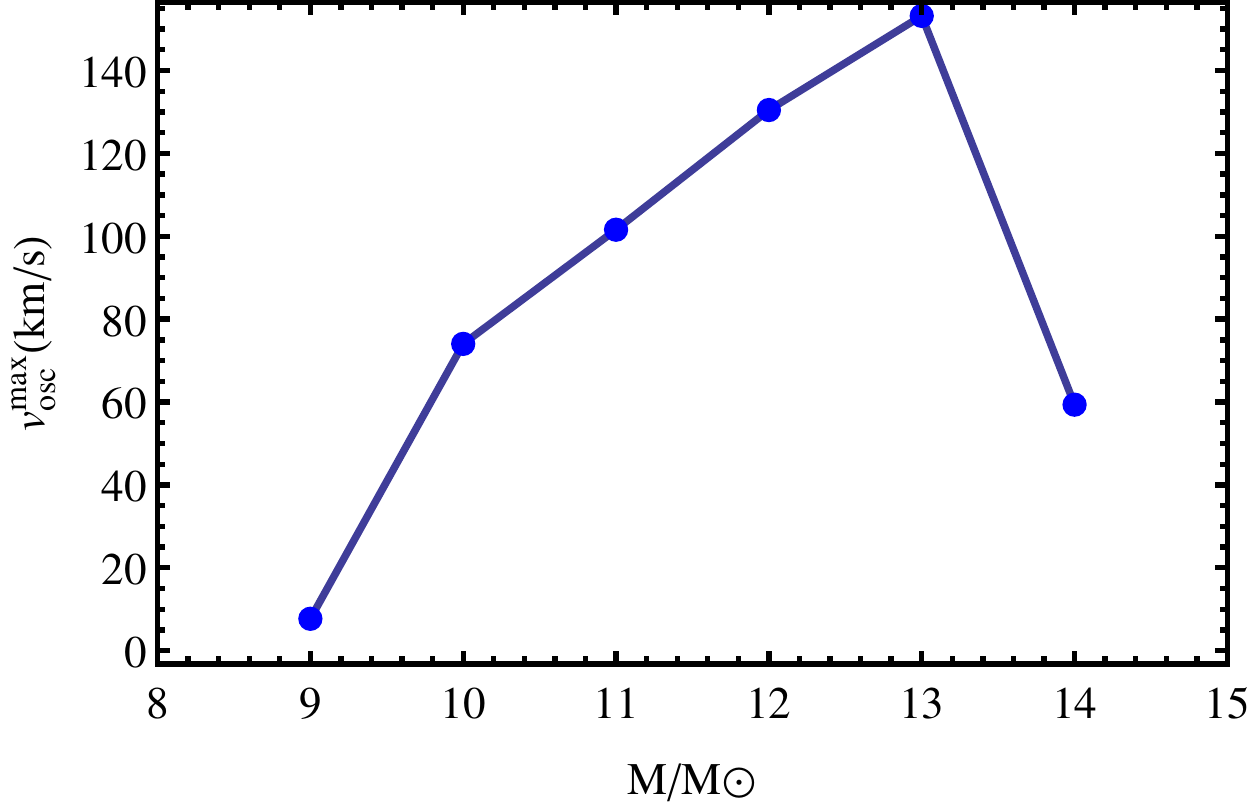}}\caption{Maximum radial velocity reached at the surface during a saturated pulsation cycle in the helium star models. The sound speed in these models is of the order of 30 km/s.
% The increase in the low mass part of the plot is connected with the growth of the extent of the inflated envelopes and the lower densities, while the decrease after 13$\rm M_\odot$ is addressed as an effect of the high mass-loss rate. Starting from the $10M_\odot$ model it is clear the supersonic nature of the strange modes, i.e. shocks are expected. 
}
\label{fig:vel}
\end{figure}

In low density extended envelopes of stars with high luminosity-to-mass ratio, such as the envelopes of our massive helium main-sequence stars (Fig.\ref{fig:logro}), the appearance of so-called strange mode pulsations is expected \citep{1993Glatzel,1998Saio,2015Owocki}. The main characteristic of these pulsations is their occurrence in regions where the ratio of the local thermal-to-dynamical timescale is small ($\tau_{th}/\tau_{dyn}\ll 1$, see also Sect.2 and Table \ref{tab2}). This is the case for WR stars and makes them good candidates for the appearence of pulsational instabilities in their radiation pressure dominated envelopes. The short thermal timescales exclude a thermal origin for these kind of pulsations, unlike  $\kappa$ and $\epsilon$ mechanisms \citep{1994Glatzel,1998Saio,2003Blaes}. 
%Strange mode pulsations are strictly connected with the opacity profile, in this case the Fe-opacity bump, and can introduce another source of instability in the envelope of WR stars \citep{1994Glatzel,1998Saio,2002GlatzelNR,2013Grafener,2015Owocki}.

 We use the same hydrodynamical stellar evolution code introduced in Sect. 2 to analyse the pulsational properties of our models. This code has already been used to investigate pulsations in the red supergiant phase by \citet{2010Yoon} and is tested against linear pulsation analysis for basic pulsational properties in \citet{1997Heger}. The code adopted here is fully implicit and therefore numerical damping may be present \citep{1970Appenzeller}. 
%A similar method has been used in \citet{2015Moriya}.

Radial pulsations in our models were found to be excited in the mass range 9-14$\rm \,M_\odot$. They show a short growth time of the order of few dynamical timescales, which is a characteristic feature of strange mode instabilities \citep[see Fig.\ref{fig:pulgro} and][]{1994Glatzel}. After the growth phase, all these models reach saturation. Stability tests with respect to radial perturbations for helium main-sequence models have been investigated in \citet{1993Glatzel}, who identified the set of unstable modes as strange modes. We find that our pulsation periods are in good agreement with those of the lowest order unstable modes in \citet{1993Glatzel}. The periods obtained with our models are plotted in Fig.\ref{fig:period} and listed in Table \ref{tab2}. 
%For a given unstable radial pulsation the dynamical time-scale is of the same order as the period. 
The pulsation periods almost linearly increases with stellar mass (Fig.\ref{fig:period}). This is related to the larger spatial extent of the envelope and increased dynamical timescale in the high mass models. No pulsations are found for models below $\rm 9\,M_\odot$ in contrast to the stability analysis performed by \citet{1993Glatzel}, where modes are excited for stellar masses as small as  $\rm 5\,M_\odot$. 
%This is probably related to the numerical damping.  

From Fig.\ref{fig:convpuls} we can see how the extent and position of the convective zone within the stellar model evolves during a pulsation cycle. The convective zone in Fig.\ref{fig:convpuls} is no longer strictly shaped by the iron opacity bump, but follows the perturbed layers where the density, and therefore the opacity, is higher. It moves periodically from deeper inside the star where the convective zone forms during the contraction of the envelope, up to the surface when the compression wave reaches the surface and forces the envelope to expand.  During a pulsation cycle part of the envelope shows significantly high values of the local Eddington factor, up to $\Gamma \approx 1.5-2$ in the contraction phase. 

The saturated amplitude of the pulsations $ \Delta R$ is plotted in Fig.\ref{fig:ampl}. The amplitude  is defined as the difference between the minimum and maximum radius in a saturated pulsation cycle. In Fig.\ref{fig:vel} we show the maximum surface radial velocity amplitude due to the radial pulsations ($v_{\rm osc}^{\rm max}$). Figure \ref{fig:ampl} shows an increase of the maximum pulsation amplitude from 9 to 13$\rm\, M_\odot$, which is connected to the higher radial velocity (see Fig.\ref{fig:vel}). In contrast, the 14$\rm \,M_\odot$ model has a lower maximum amplitude and maximum velocity than lower mass models. No other pulsating model has been found for masses $\rm M > 14\, M_\odot$, when mass loss was applied. Conversely, we find that higher mass models without mass loss are also unstable to more than one mode. We therefore interpret the decrease of the amplitude for the models with higher masses as the result of the higher mass-loss rate. 
For the pulsating 14$\rm\, M_\odot$ model, which shows $ \Delta R/R$ of the order of 0.03 with no mass loss applied, the pulsation amplitude is smaller when mass loss is applied ($ \Delta R/R=0.0085$). In these cases the amount of mass depleted from the inflated envelope during a pulsation cycle becomes larger than 5\%, which is equivalent to 40\% of the envelope radial extent in a hydrostatic non-pulsating configuration. 

Although the WR mass-loss rates proposed by \citet{2000Nugis} are approximately 2 orders of magnitude smaller than the critical mass-loss rate necessary to remove the inflated envelope \citep{2006Petrovic}, the absence of radial pulsations for masses $\rm M>14\,M_\odot$ can be explained as a stabilizing influence of mass loss on the strange modes instability. These models are in fact expected to be unstable in the context of the analysis carried out by \citet{1993Glatzel}. However, neither \citet{1993Glatzel} nor any other author (at the best of our knowledge) included mass loss in their stability analysis.
%["Glatzel talks about the stabilizing influence of stationary winds on the strange modes instability in \citet{1993Glatzel} sect.conclusion"]. 

From Fig.\ref{fig:vel} we notice that pulsations occur with supersonic velocities with values up to a factor 4 higher than the isothermal sound speed. This is another peculiar feature of the strange mode instability \citep{1994Glatzel,2005Glatzel}. Instead, the luminosity variations during a saturated pulsation cycle are very small (see Tab.\ref{tab2}), corresponding to a variability of the order of $10^{-7}$ bolometric magnitudes.
The results obtained here show that strange modes in mass-losing, H-free WR stars occur in the mass range 9-14$\rm M_\odot$ with periods of the order of minutes, amplitude of the order of 1\% of the stellar radius, and supersonic radial velocities. The amplitude of the pulsations increases as a function of mass in the range 9-13$\rm\, M_\odot$, while pulsations are damped or inhibited for the more massive models considered here, i.e. $\rm M \gtrsim 13\,M_\odot$ and with $ \rm \dot{M}  \gtrsim -10^{-5} \,M_\odot/yr$. Consequently, the most massive H-free WR stars may not be unstable against pulsations. Therefore, the role of pulsations in driving the wind or in inducing instabilities in WR stars may not be crucial, or their importance can be restricted to the mass range 9-14 $\rm \,M_\odot$. 

\section{Comparison with observations}

\begin{figure}
\resizebox{\hsize}{!}{\includegraphics{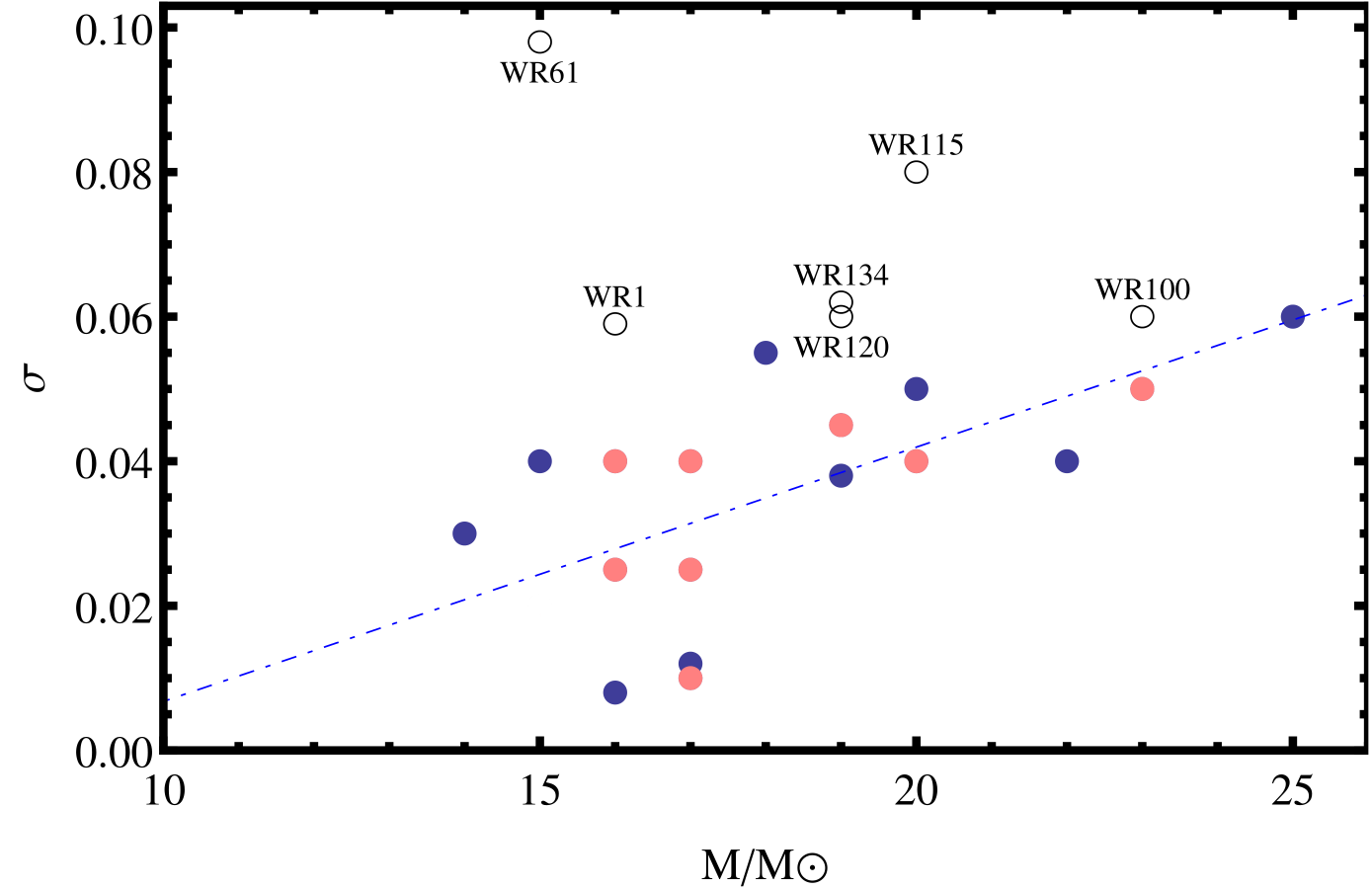}}\caption{ Variability of rms relative to the line strength $\sigma$ as a function of mass of the Galactic H-free WN stars. The blue dots correspond to the small-scale variability in WN stars from \citet{2009StLouis,2011CheneStLouis,2014Michaux}, while the pink dots correspond to the variability for the stars analysed in this work (see Appendix A). Empty black circles correspond to CIR-type variability \citep{2009StLouis,2011CheneStLouis,2013StLouis}. The stellar masses are derived by \citet{2006Hamann}. The linear fit in blue dashed is computed considering only the blue and pink dots. The Spearman's rank correlation coefficient is $\approx 0.7$.}
\label{fig:MVar}
\end{figure}

One of the dominant features characterizing WR stars is their strong, partly optically thick wind. Such a feature shrouds the hydrostatic surface of the star and we are only able to have indirect evidence of the physical conditions at the base of the wind.

\subsection{Spectral variability}
 
Seeking to test the influence of the velocity field generated by the sub-surface convective zone and to relate it to the formation of clumps, we compare our results with the recent studies by \citet{2014Michaux}, \citet{2011CheneStLouis}, and \citet{2009StLouis}. These works focus on WR stars and the variability of WR spectral lines and are based on atmosphere models from \citet{2012Sander} and \citet{2006Hamann}. Additionally, we analyse the spectra of eight WR stars of the WN spectral type as described in Appendix A.

Our goal is to verify whether an increasing small-scale variability level as a function of the estimated mass can be found in some spectral lines, as expected from Fig.\ref{surfconv}. Following \cite{2009StLouis}, \cite{2011CheneStLouis}, \cite{2014Michaux}, and the formalism of \cite{1996Fullerton}, we consider the rms variation across the He{\sc ii} wind spectral lines relative to the local continuum strength identified as variable $\sigma$. We did this to investigate the spectroscopic variability in the sub-sample of H-free WR stars of the WN subtype \citep[see][and reference therein for further details]{2011CheneStLouis}. The degrees of spectral variability $\sigma$ were derived in general from only four to five spectra per star, which does not allow a direct distinction between stochastic and periodic variability. This in turn does not always allow for a direct distinction between WR stars showing CIRs and WR stars with only clumping, which is achieved by characterizing the kinematics of the excess emission on top of the emission lines that either move from one line edge to the other in the case of CIRs, or from the line centre to the line edges in the case of clumps.   

From \citet{2014Michaux}, which is mostly a compilation of the data of \citet{2009StLouis} and \citet{2011CheneStLouis} for the WN class, we exclude the objects showing to some extent hydrogen lines. In fact, H-rich WN stars can be structurally different from the H-free models that have been computed in this work. Additionally, we exclude stars that have been reported to most likely have CIR-type variability \citep{2009StLouis,2011CheneStLouis,2013StLouis}. The CIR-type variable spectra are associated with large-scale variability, which most likely does not find its origin in the small-scale inhomogeneity investigated here. Therefore these objects are not included in the linear fit in Fig.\ref{fig:MVar} (empty circles) and in the following analysis. In other words, only the sample of single\footnote{Some WR stars might have an O-B companion, which in any case does not significantly affect the WR spectrum and derived physical quantities, according to \citet{2006Hamann}. } H-free WN stars has been plotted in Fig.\ref{fig:MVar} as a function of the estimated mass.

Most of the spectral variability studies on which the $\sigma$ values are extracted are presented in the works of \citet{2009StLouis} and \citet{2011CheneStLouis}. An additional eight WN stars, however, partially presented in \citet{2012Chene} were also included (pink dots in Fig.\ref{fig:MVar}). They are the Galactic stars WR\,7 (WN4b), WR\,20 (WN5O), WR\,34 (WN5o), WR\,37 (WN4b), WR\,51 (WN4o), WR\,62 (WN6b), WR\,84 (WN7o), and WR\,91 (WN7b). Each of these stars was observed four to eight times with the Gemini Multi-Object Spectrograph (GMOS) at Gemini South, under programme numbers GS-2008B-Q-87, GS-2010B-Q-58, GS-2014A-Q-42, and GS-2014A-Q-73. In Appendix A, we present the spectra and a detailed variability analysis comparable to what is available in the literature for other WN stars.

The observed WR stars only partially fall in the mass range presented in Fig. \ref{surfconv}. However, the linear fit in Fig.\ref{fig:MVar} shows larger variability at higher stellar masses. This trend is in agreement with the results shown in Fig.\ref{surfconv}. In addition the zero point of the linear fit at $\rm 10\, M_\odot$ matches our prediction. Moreover, we do not find any correlation when the degrees of spectral variability are compared to the terminal wind velocities tabulated by \citet{2006Hamann}, suggesting that the spectral variability is not related to the specific wind structure but rather by the conditions in the deeper hydrostatic structure of the WR stars. 

 While these results have to be taken with caution given the small sample and the associated uncertainties, we conclude that our results are supporting the idea that sub-surface convection may trigger the formation of clumps in the wind of Wolf-Rayet stars.
%Moreover, following \citet{2006Petrovic} and Fig.~\ref{fig:HRD} a bending of the He-ZAMS is seen at luminosities above $\rm Log(L/L_\odot)\geq 5.5$ due to the inflated envelope, which would imply an anti-correlation between mass and temperature, in agreement with the results in \citet{2014Michaux} and the results of this analysis. \\ 

%\textcolor{red}{Interesting is also the fact that in the data from \citet{2014Michaux} one can see a high variability in the effective temperature range of the so called He-opacity bump ($\rm log(T) \approx 4.7$) and the relative convective zone. In the case of WNE stars the He-opacity bump is expected to be in the highly supersonic part of the wind, being at lower temperature respect to the surface temperatures of these models. In the case of WNL stars instead the He-opacity bump is in the proximity of the location of the sonic point and of the photosphere. Therefore, the variability reported in \citet{2014Michaux} in the temperature range of the He-bump could eventually be connected to the joined effect of perturbations with an origin in both the iron bump CZ deeper inside the star and the helium bump CZ in the proximity of the surface. }  

\subsection{Characteristic timescale and number of clumps}

A further comparison with observational data could involve the characteristic timescale associated with convection, $\rm \tau_{conv}$, defined as
\begin{equation}
\tau_{conv}=\frac{\alpha H_P (R_c)}{\mean{v_c}}\quad.
\label{timeconv}
\end{equation}
This timescale is of the order of $10^2-10^3$ seconds in our models with $\rm M\geq10 \,M_\odot$ (Table \ref{tab1}). Clumping could be detected by means of linear polarimetry \citep[see][]{2005Davies,2007Davies,2012Vink,2013StLouis} and a Fourier analysis of its variability could lead to the direct detection of the sub-surface convection in WR stars where the optical depth of the wind is sufficiently low. 

The derived timescale appears to be in good agreement with spectroscopic variations associated with small-scale structures in the wind of WR stars. \citet{1999Lepine3} obtained a time series of high-resolution and high signal-to-noise ratio spectra of the WC8 star $\gamma ^2$ Velorum with a relatively high sampling rate of one spectrum every 300s. From these observations, they concluded that the temporal variability of the sub-peaks associated with clumps in the wind took place on a timescale of the order of minutes to hours \footnote{The variability timescale does not have to be confused with the time necessary for a sub-peak to migrate from the line centre to the line edges.}.

The timescale might also be observable by detecting short time variations in different parameters of linear polarimetry such as the total linear polarization or the angle of maximum polarization. The detection of the stochastic variability on such a short timescale requires the use of extremely short exposures and readout times, therefore limiting both the number of usable instruments and observable targets. In this respect it might be best to use instruments capable of obtaining both linear polarimetry Stokes parameters $Q$ and $U$ simultaneously to halve the observing time.

 Furthermore, we can compare the estimated order of magnitude for the number of clumps as from Eq.\ref{nclumps}, i.e. about $10^3-10^4$ clumps, with observations. The wind crossing time, usually of the order of hours \citep{1996Lepine,2008Chenetime}, is comparable to the excitation timescale above, therefore even if our estimate shall be considered as a lower limit, we do not expect this to differ significantly compared to the integrated number of clumps throughout the stellar outflow.
%when we compare the estimated order of magnitude for the number of clumps as from Eq.\ref{nclumps}, i.e. about $10^3-10^4$ clumps, with observations, 
We find that indeed our estimate is in agreement with the number of clumps observed by \citet{1999Lepine} for WR stars ($\approx 10^4$) as well as with \citet{2007Davies} for O and LBV stars. 

\subsection{Magnetic fields}

The formation of magnetic flux tubes could also be connected to sub-surface convection in hot rotating stars \citep{2002Prinja,2003MacGregor,2009Cantiello,2010Cantiello,2011Cantiello}. A magnetic field can be generated from a dynamo mechanism taking place in the FeCZ and propagate through the outer layers of WR stars, affecting their winds and eventually the structures in these winds. The presence of extended FeCZs in the most massive WR models could be associated with different phenomena as was successfully accomplished in the case of O-stars \citep[e.g.][]{1995Prinja,1998Prinja,2011Cantiello}. These phenomena are discrete absorption components \citep[e.g.][]{1996Cranmer}, i.e. X-ray emission \citep[e.g. ][]{2003Kramer,2009Oskinova} and large-scale variability induced by corotating interaction regions \citep[e.g.][]{1996Cranmer,2011CheneStLouis,2011CheneCIR}, which appear in several WR star winds. 

Based on the model for the rise of toroidal magnetic flux tubes generated by a dynamo mechanism in the convective zone proposed by \citet{2003MacGregor}, who investigated the buoyant transport of magnetic flux through the interior of hot stars, we estimate the field strength of the magnetic flux tubes in the sub-surface convective zone. Assuming equipartition between the magnetic and kinetic turbulent energy as in \citet{2009Cantiello}, we can write the magnetic field strength {\it B} as
\begin{equation}\label{magnetic}
B \simeq \mean{v_c} \sqrt{4 \pi \rho_c} \quad.
\end{equation}

From Eq.\ref{magnetic} we derive a magnetic field at the interface between the convective and overlying radiative zone of the order of tens of Gauss for the low-mass ($\lesssim 10\Msun$) helium
stars , while for the more massive models we find magnetic fields of up to 500 G (cf. Tab.\ref{tab1}). If we assume conservation of the magnetic flux for the magnetic flux tubes \citep{2003MacGregor}, we find that the strength of the magnetic field arising at the surface is roughly the same as that at the interface between the convective and radiative layers, i.e. of the order of hundreds of Gauss in the models with $M \gtrsim 10 \Msun$.   
However the magnetic pressure at the surface is still only a fraction of the surface gas pressure, i.e. at maximum a third of it in the most massive model. This implies that the gas and radiation pressure most likely dominate and define the global structure of the outflow at the base of the wind \citep{2001Gary}, but local twisting of the magnetic field lines might lead to locally enhanced magnetic field strength such that the magnetic pressure may exceed the local gas pressure with potential observational effects, such as CIRs.     

 Despite the simplified method and assumptions, and the observational difficulties in detecting and estimating the magnetic fields of WRs, we find that the predicted values are of the same order of magnitude as the most probable values determined for a sub-sample of WR stars by \citet{2013DelaChev,2014DelaChev}. 
 These are marginal detections of the field in the observable region of the wind and that the field configuration was assumed to have a split monopole configuration.

 \subsection{Pulsations}

%, show higher spectral variability at lower surface temperature, which may appear to be in contrast with Fig.~\ref{surfconv} .\\ However the comparison is not straight forward first because H-rich WR stars are included in the studied sample and H-rich models can be . Second the feedback of the optically thick wind is not taken into account in this work, which might influence the scenario here reported, as already stated \citep{2002Nugis}. Third the theoretically reasonable assumption in \citet{2014Michaux} of a correlation between mass and temperature happens to be not correct in the specific case of the H-free WN stars in the sample. In fact considering only H-free WN an anti-correlation between mass and temperature can be found. 

It has been suggested that violent mass-loss events and large-scale variabilities may be connected with strange mode instabilities in hot massive stars \citep{1993Glatzel,2002GlatzelNR}. %In this vein pulsations should get rid of part or all the envelope of the star due to the reduced effective gravity at the surface and the supersonic pulsation velocities, and consequently trigger preferably large scale variability in the spectra of WR stars especially considering the non-radial nature of the strange modes \citep{1993Glatzel,2002GlatzelNR}.
A clear observational signature for fast pulsating WR stars has not yet been found. After the non-confirmation of the presence of fast pulsations in WR40 of spectral sub-type WN8 \citep{1992Blecha,1994Martinez}, the only candidate at the moment is the H-depleted WR123 of the spectral sub-type WN8, which shows a 9.8 hours period. This period however does not match the period range of the lowest order unstable modes in Fig. \ref{fig:period}  \citep{2010Balona,2005Lefevre,2011CheneWR123}. However, WR123 is expected to be considerably more massive then the mass range considered here. Previous attempts to reproduce this 9.8h period have been unsuccessful \citep[see][and reference therein]{2011CheneWR123}.

  The predicted bolometric magnitude change of our models during a pulsation cycle is of the order of $\Delta m \propto log(L_{min}/L_{max})\approx 10^{-7}$ (see Tab.\ref{tab2}). However, the associated changes in effective temperature imply a variability as high as 0.1\,mag in the X-ray and visible bands. While this may be above the level of non-periodical variability of WRs at the mmag level \citep{2009Gosset,2010Balona,2011CheneWR,2012Uraz}, a periodic variability in the frequency range we predict  has not been detected in the bright Galactic WC5 star WR111 by \citet{2008MoffatCoherence} \citep[see also ][]{2005Lefevre,2011CheneCIR}. The pulsations may be generally hidden by other types of variability appearing in WR outflows and by the optically thick winds, given the high optical depth of the hydrostatic layers. Whether any variability induced by the pulsations is detectable depends on how pulsations affect the stellar outflow and whether coherence is preserved throughout the wind.
From Fig.\ref{fig:ampl} and the discussion in Sect.4, it is suggested that the wind has a stabilizing effect on strange modes instability and inhibits them in most cases. This might explain why, even if expected, there has been no match between the theoretical predictions and observations of strange modes pulsations in H-free WR stars, as previous models neglected the effect of mass loss to the stability of strange modes.

A different explanation for small- or large-scale variability might instead involve the effects of turbulent pressure fluctuations and finite amplitude high order non-radial pulsations of non-thermal origin. These are expected to appear in the case of transonic convective velocities and inefficient convection, as has been suggested to connect the strength of turbulent pressure in the partial ionization zones to the appearence of the macroturbulent broadening in OB and late-type stars; this is a broadening of  spectral lines on a scale larger than the line forming region \citep{2015GrassitelliB,2015GrassitelliA}. In our models the turbulent pressure contributes less then 1\% to the equation of state, but the Mach numbers are close to unity in the more massive models.

\section{Conclusions}

The purpose of this work was to study the condition in the envelopes of massive helium stars and analyse the instabilities appearing in these radiation pressure dominated regions. These models correspond to H-free Wolf-Rayet stars, which are located close to the Eddington limit. These instabilities can possibly provoke or influence the formation of structures observed at the surface or in the wind.
We find that sub-surface convective zones are present in all the investigated models. Their spatial extent increases for higher stellar masses. The convective zones arise as a result of the iron opacity peak and they are inefficient in transporting energy. Convection reaches the surface in the $\rm 17\,M_{\odot}$ model.
In our models the iron opacity bump is also responsible for the envelope inflation \citep[][]{2015Sanyal} starting from the $\rm 6\,M_{\odot}$ model. These envelopes are characterized by their proximity to the local Eddingont limit (i.e. $\rm \Gamma \approx 1$). 

The convective motion may trigger velocity and density fluctuations at the stellar surface.  An upper border for the surface turbulent velocity is estimated by considering the energy transported via gravity waves from the upper limit of the sub-surface convective zone to the surface. We find that the expected surface velocity amplitudes are small ($\rm \lesssim 1\, km/s$) for stellar models with masses below $\rm M \lesssim 10 \, M_\odot$, but are more than 20 km/s for the more massive stellar models  ($\rm M \gtrsim 10\, M_\odot$). Therefore if clumping is triggered at the base of the wind by these fluctuations, we expect that the formation of structures should be inhibited or spatially delayed \citep{1995Gayley,2015Owocki} in the case of low-mass WNE stars. However, our models predict strong small-scale variability for higher WR masses. 
A trend of stronger variability for higher stellar masses was identified from the observational data of \citet{2009StLouis}, \citet{2011CheneStLouis}, and from this work (see Sect.5.1 and Appendix A) for single H-free WNE stars. We also conclude that if wind clumping is triggered by sub-surface convection, the assumption of a constant clumping factor as a function of mass may not be appropriate. We find also that the timescales associated with convection, of the order of hours, are supported by observations by \citet{1999Lepine3}. Moreover, the number of clumps that we estimate by considering the size of the biggest convective eddies and the correlation lengths associated with the gravity waves, i.e. $10^3-10^4$, is also supported by observations by \citet{1999Lepine}. 

%We point out that our analysis does not take into account the propagation of the waves in the supersonic optically thick winds, nor a complete study on the dissipative effects and the trans-sonic behaviour of the waves. Consequently, a precise observational diagnostic can hardly be explicitly predicted. One as to be aware of the possible strong influence of the optically thick winds on the structure of the envelope of the helium star models, e.g. backwarming effect, different dynamic, etc. \citep{1999LamersCassinelli,2002Nugis,2013Grafener,2015Owocki}.
%, which is the main limit of this work and will be object of future works from the authors. \citep{2013Sundqvist}.

We find that the envelopes of the helium star models are unstable to pulsations in the range $\rm 9-14\, M_\odot$. The periods of the pulsations are in agreement with a low order mode in \citet{1993Glatzel}. The variations in bolometric magnitude connected with these pulsations are very small (less than micromagnitudes), i.e. below the observable photometric variability and the observed variabilities connected with structures in the wind. However the supersonic radial velocities could give rise to observable periodical effects  apparently not yet observed in the context of WR stars. 
%{\bf Moreover, the corresponding variations in magnitude for the continuum at the wavelength of the He{\sc ii}$\lambda$5411 spectral line can be as high as 0.1, suggesting that this kind of pulsations, if present and not affected by the optically thick layers, should be easily detectable.}   
Moreover, the strong mass loss applied to our stellar models has a stabilizing effect on the pulsations. It appears that mass loss has an inhibiting effect on the pulsational instability in the most massive cases. This may help to explain the lack of observational evidence for periodical variability in WR stars of the WNE type, which is in contrast to previous direct theoretical predictions that did not include mass loss.  

We also find that the strength of the magnetic fields rising from the convective zone to the surface of our most massive models, of the order of hundreds of Gauss, is comparable to the most probable values determined in some WR stars by \citet{2013DelaChev,2014DelaChev}. The pressure induced by the buoyant magnetic flux tubes is, however, globally not enough to define the structure of the wind, but local enhancement of the magnetic field  might lead to or trigger the formation of CIRs.

\begin{acknowledgements} L.G. is part of the International Max Planck Research School (IMPRS), Max-Planck-Institut f{\"u}r Radioastronomie and University of Bonn and Cologne. L.F. acknowledges financial support from the Alexander von Humboldt foundation. Based on observations obtained at the Gemini Observatory, processed using the Gemini IRAF package, which is operated by the Association of
Universities for Research in Astronomy, Inc., under a cooperative agreement
with the NSF on behalf of the Gemini partnership: the National Science
Foundation (United States), the National Research Council (Canada), CONICYT
(Chile), the Australian Research Council (Australia), Minist\'{e}rio da
Ci\^{e}ncia, Tecnologia e Inova\c{c}\~{a}o (Brazil) and Ministerio de
Ciencia, Tecnolog\'{i}a e Innovaci\'{o}n Productiva (Argentina). Further, L.G. thanks T.Moffat, Y.Michaux, G.Gr{\"a}fener, J.-C.Passy, and the referee S.P.Owocki for precious discussions and comments on this manuscript. 
\end{acknowledgements}

\bibliographystyle{aa}
\bibliography{conv}

\appendix

\section[]{Variability study of eight additional WN stars}

\subsection[]{Data extraction and analysis}

The bias subtraction, flat fielding, spectrum extraction, sky subtraction, and wavelength calibration of all spectra were executed in the usual way using the gemini packages of the {\sc iraf}\footnote{{\sc iraf} is distributed by the National Optical Astronomy Observatories (NOAO), which is operated by the Association of Universities for Research in Astronomy, Inc. (AURA) under cooperative agreement with the National Science Foundation (NSF).} software. Calibration lamp spectra were taken during the night after each spectrum. Special care was taken for the normalization of the spectra. First, a mean was made for each run. Then, each spectrum of a given star was divided by the mean spectrum and the ratio fitted with a low-order Legendre polynomial (between fourth and eighth order). The original individual spectrum was divided by this fit, and was therefore at the same level as the mean spectrum. This allowed us to put all individual spectra at the same level. The mean spectrum was then fitted in selected pseudo-continuum regions, i.e. wavelength regions where large emission lines do not dominate. Finally, the fitted continuum function is applied to each individual spectrum. The error on the continuum normalization measured as the standard deviation of individual spectra around the continuum function is typically of 0.5%. To ensure that the detected variability would not be affected by any motion in radial velocity (or wavelength calibration uncertainty), we have aligned the spectra using cross-correlation. We have also corrected the line intensity by forcing a constant equivalent width through the time series, to avoid any variation in the continuum level to cause variable line dilution that could hide the intrinsic wind variability of interest.

\subsection[]{Determining the $\sigma$-value}

The final spectra are shown in the top panels of Figures A1-A2 for the three strongest lines present in our wavelength interval, i.e. the He{\sc ii}$\lambda$4686, He{\sc ii}$\lambda$5411, and He{\sc i}$\lambda$5876. 

To determine the level of line-profile variability, we first identify at which wavelengths the spectrum may be significantly variable, using the temporal variance spectrum (TVS) of each data set. The TVS was calculated using the formalism of \citet{1996Fullerton} and the quantity $\Sigma_j(99\%)=\sqrt(\frac{(TVS)_j}{\sigma_0^2 \chi_{N-1}^2 (99\%)})$ was obtained, where $\sigma_0$ is the reciprocal of the rms of the noise level in the continuum in a time series of N spectra. The value of $\Sigma_j(99\%)$ quantifies the level of variability at each wavelength: a spectrum that reaches a value of n varies with an amplitude $n$ times higher than the variability measured in the continuum (which is assumed to be pure noise) with a confidence level of 99\%. The spectrum of a given star is 
considered significantly variable at a given wavelength $j$ if the value of $\Sigma_j(99\%)$ is significantly greater than 1. When a line is identified as significantly variable, it is possible to calculate the amplitude of its variability relative to its intensity. To accomplish this, we calculated for each wavelength $j$ a modified $TVS^{1/2}_j$ as defined in \citet{2007Chene} and divided it by the line flux ($\bar{S}_j -1$), where $\bar{S}_j$ is the weighted mean flux at wavelength $j$. This ratio is named the $\sigma$ spectrum. One should note that the calculation of $\sigma$ does not take into account instrumental variations due to the noise level. Thus, when the variation level of a given line is too close to the noise, which is the case for weaker lines, $\sigma$ is artificially high. That is why we manually set $\sigma_j = 0$ when the variability at a given wavelength $j$ is not clearly significant according to the value of $\Sigma_j (99\%)$ or when the line has a relative intensity that is lower than 1.5, which is the intensity limit of a spectral line from which we can investigate the variability for the present data set. 

Figures A1 and A2 show our results for the eight additional WN stars. The bottom panels show a superposition of all the observed spectra with the average spectrum overplotted in red. The line intensity is normalized to ease the comparison with the different lines. The normalizing factor of the lines is written on the top left. The second panels from the bottom show the $\Sigma (99\%)$ spectrum. The dotted line marks the threshold of significant variability. The third panels from the bottom show the $\sigma$ spectrum. The most reliable values are extracted from the central region of the line, as the edges tend to be more sensitive to instrumental effects. Finally, the top panels show the montage of the residuals (individual spectra - average). The residuals of the He{\sc ii} $\lambda$ 4686 line are divided by a factor 2 for a better visibility. The residuals are in chronological order, with the bottom to the top. The heliocentric Julian date corrected to start on 1 Jan 2000 is written on the far left of the plot. All the spectra and associated values ($\Sigma (99\%), \sigma$) are masked  at the interstellar medium Na I D lines at 5885 and 5890\AA.

\begin{figure*}
\begin{subfigure}{.4\textwidth}
%\centering
\includegraphics[width=9cm,height=11cm]{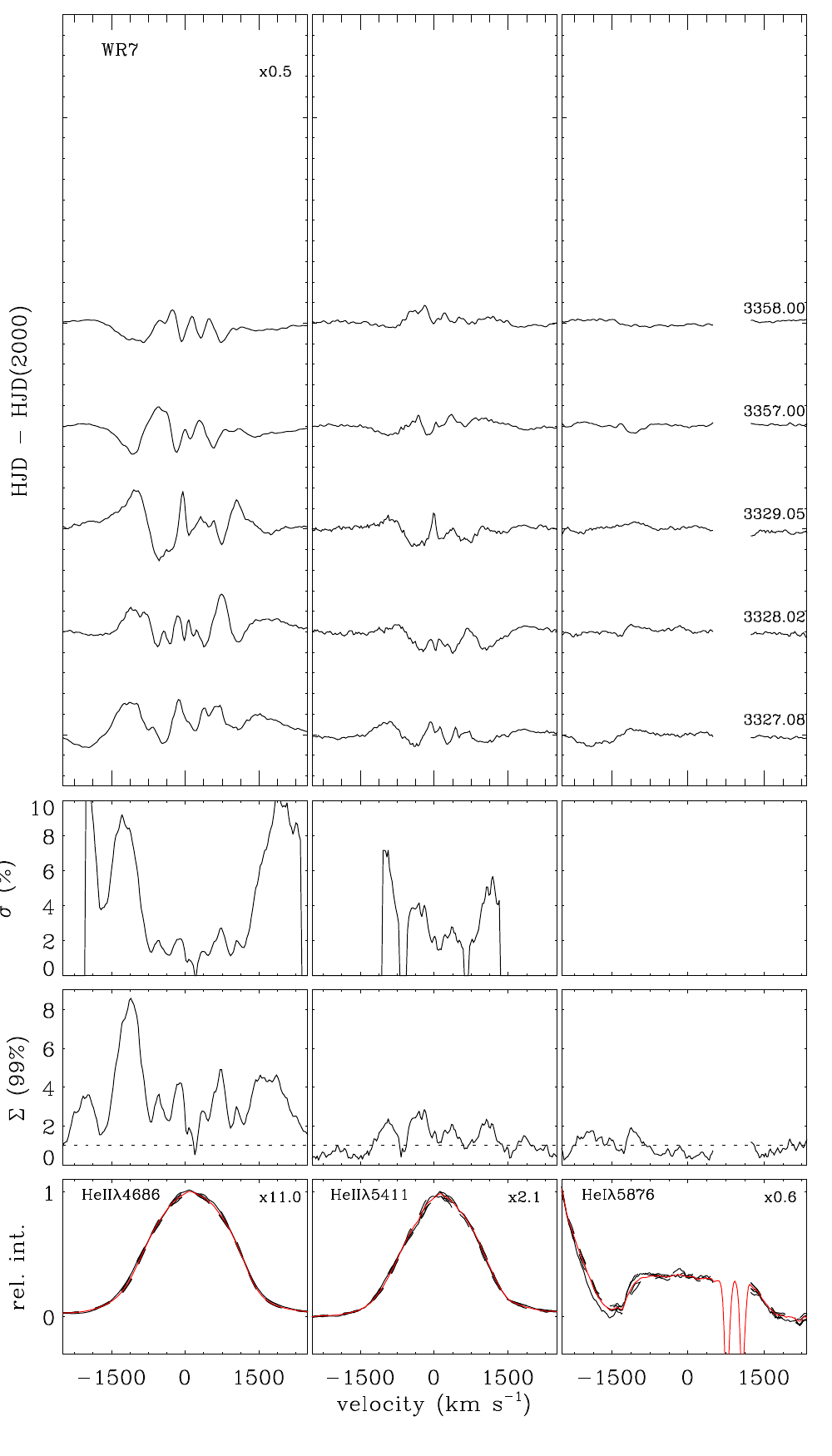}
\end{subfigure}
\begin{subfigure}{.4\textwidth}
%\centering
\hspace{-1.5cm}
\includegraphics[width=9cm,height=11cm]{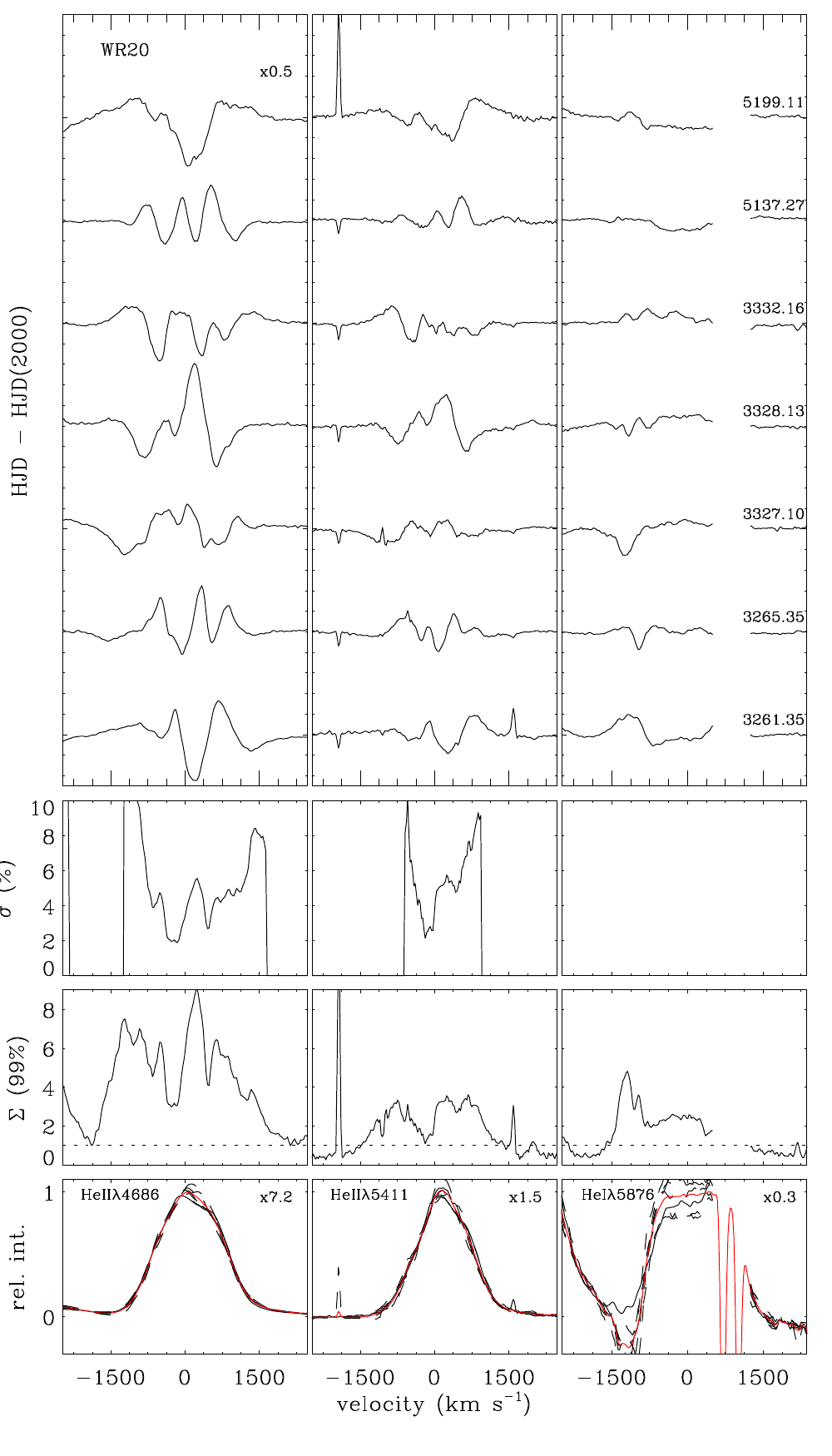}
\end{subfigure}
\begin{subfigure}{.4\textwidth}
%\centering
\includegraphics[width=9cm,height=11cm]{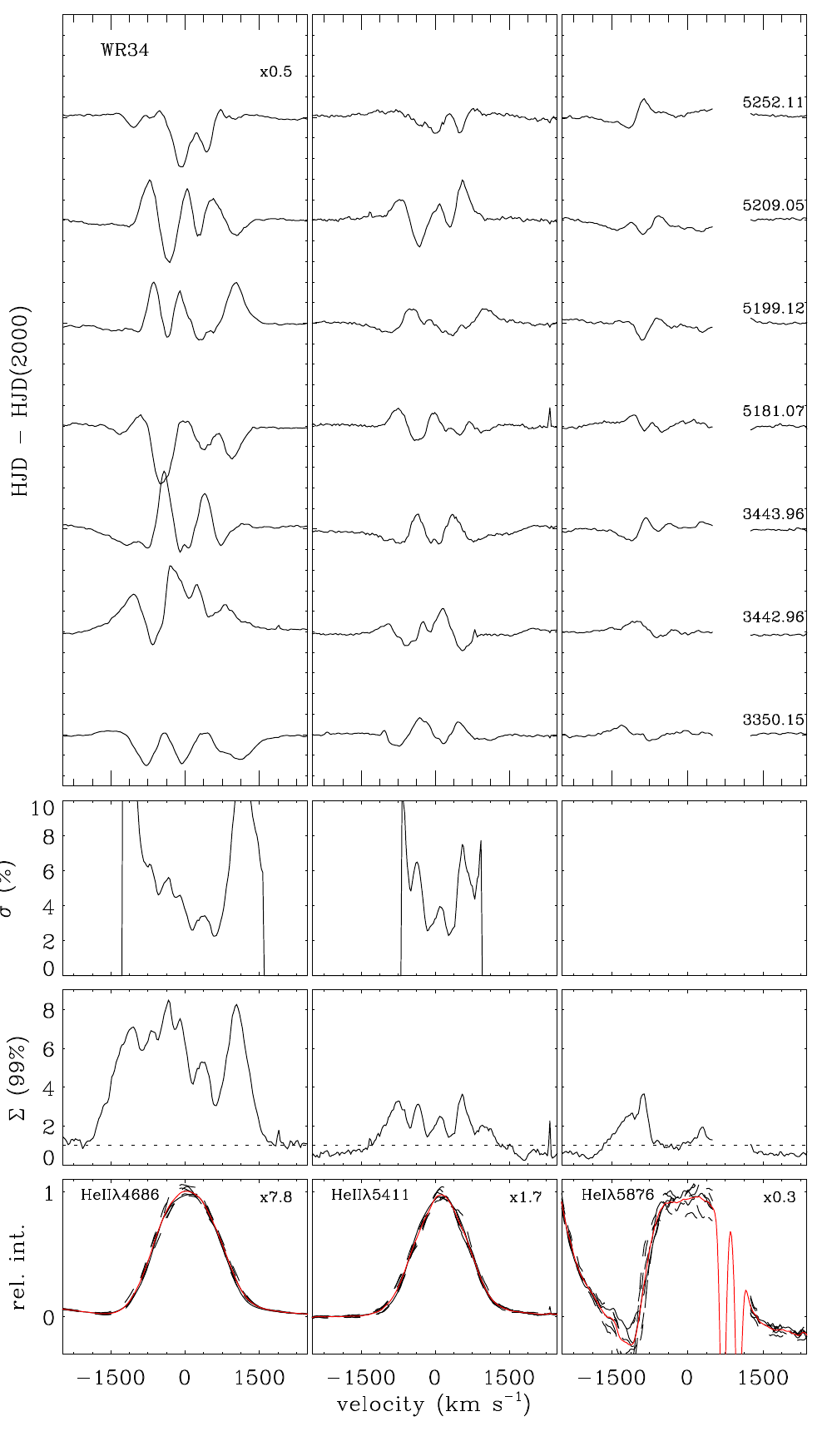}
\end{subfigure}
\begin{subfigure}{.4\textwidth}
%\centering
\hspace{2.1cm}
\includegraphics[width=9cm,height=11cm]{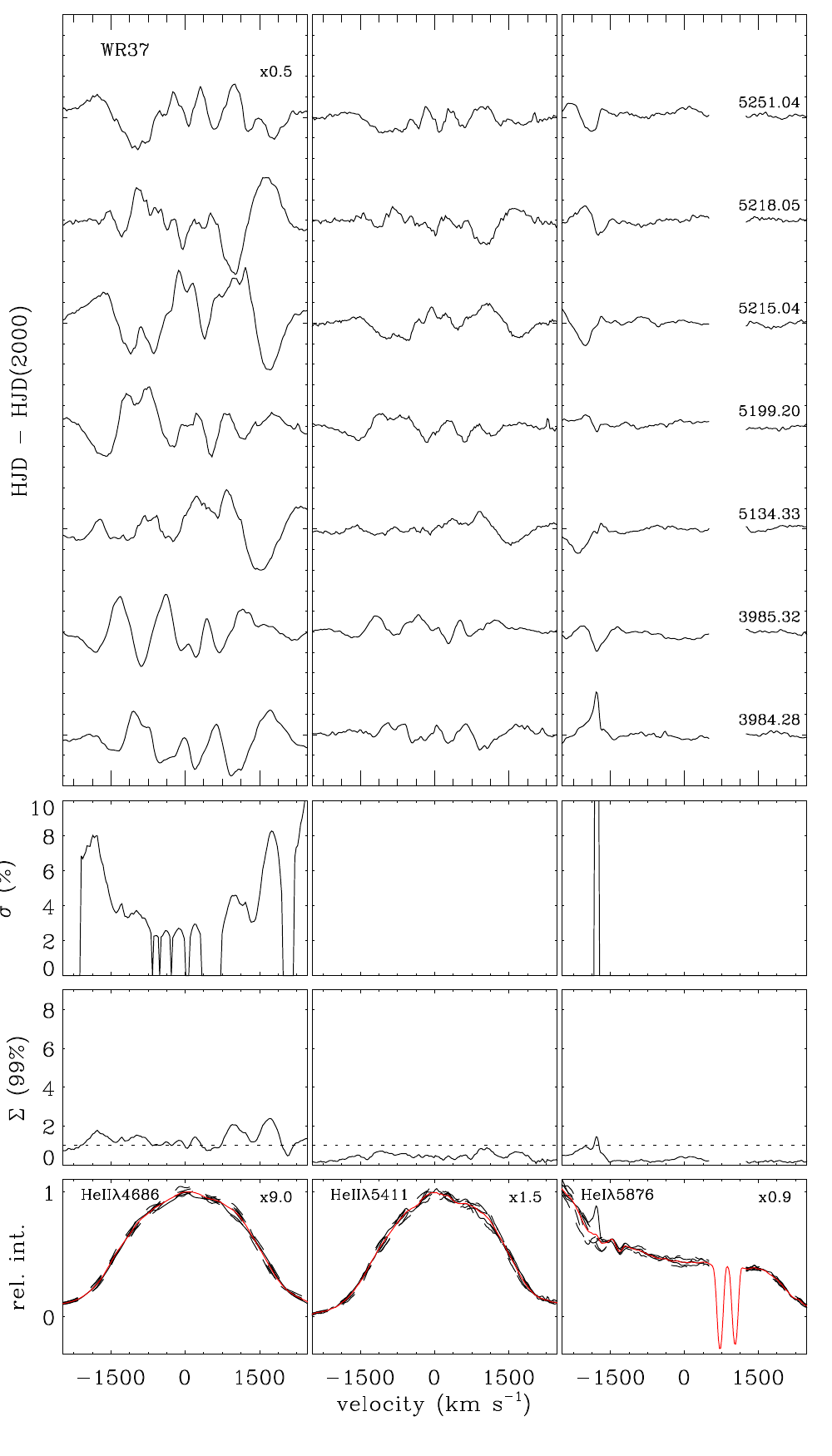}
\end{subfigure}

\caption{ a)top: Montage of the He{\sc ii}$\lambda$4686 (left), He{\sc ii}$\lambda$5411 (centre), and He{\sc i}$\lambda$5876 (right) residuals (individual spectra mean) for WR7, WR20, WR34, and WR37 (as indicated in the top left section of the top panel). For all cases, the scale factor of the ordinate is indicated in the top right-hand corner of the left column. HJD - HJD(2000) is indicated in the y-axis. Second from top: $\sigma$ spectrum. Second from bottom: $\Sigma$ (99$\%$) spectrum. Bottom: Mean spectrum.  
}
\end{figure*}

\begin{figure*}
\begin{subfigure}{.4\textwidth}
%\centering
\includegraphics[width=9cm,height=11cm]{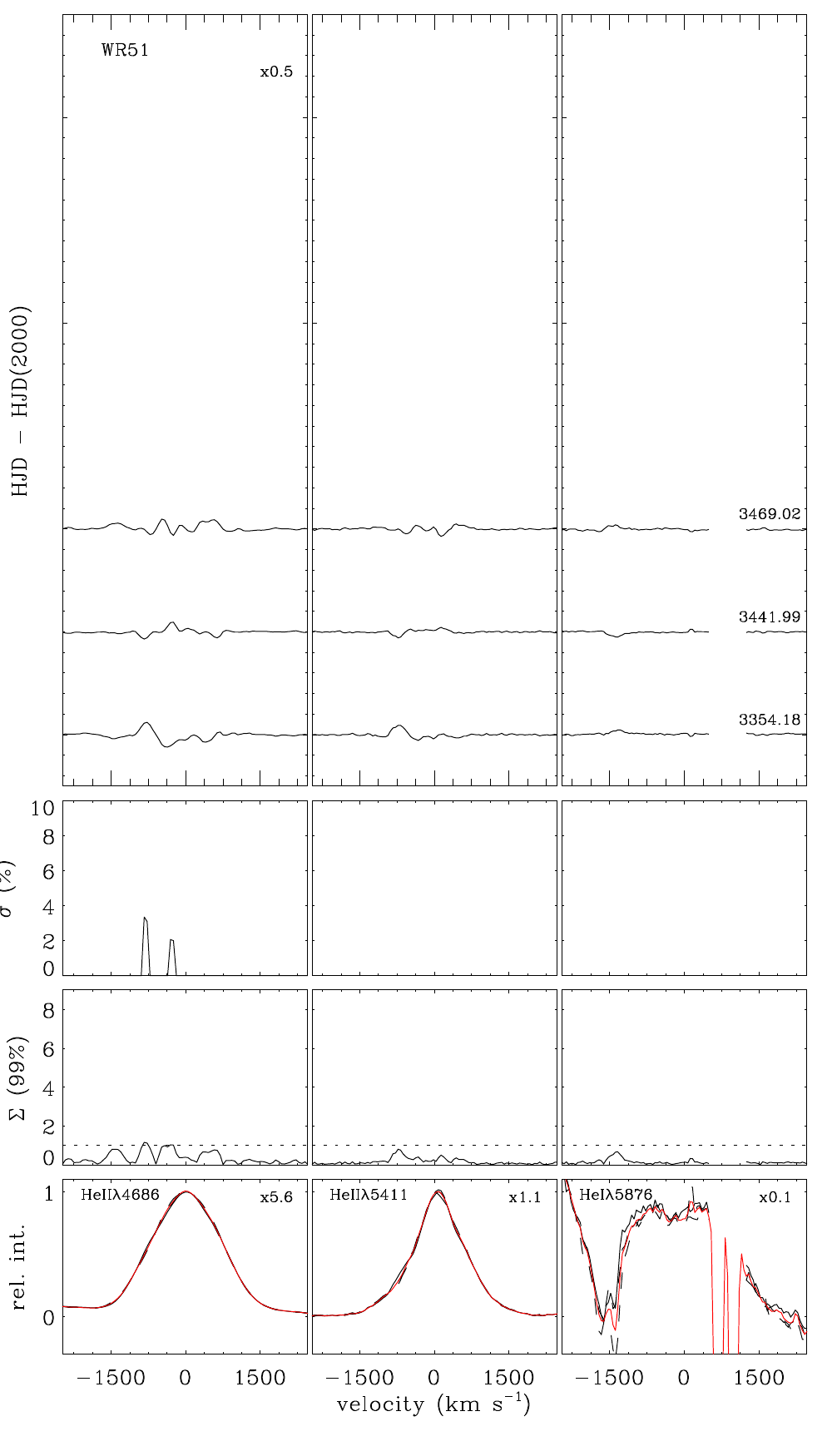}
\end{subfigure}
\begin{subfigure}{.4\textwidth}
%\centering
\hspace{-1.5cm}
\includegraphics[width=9cm,height=11cm]{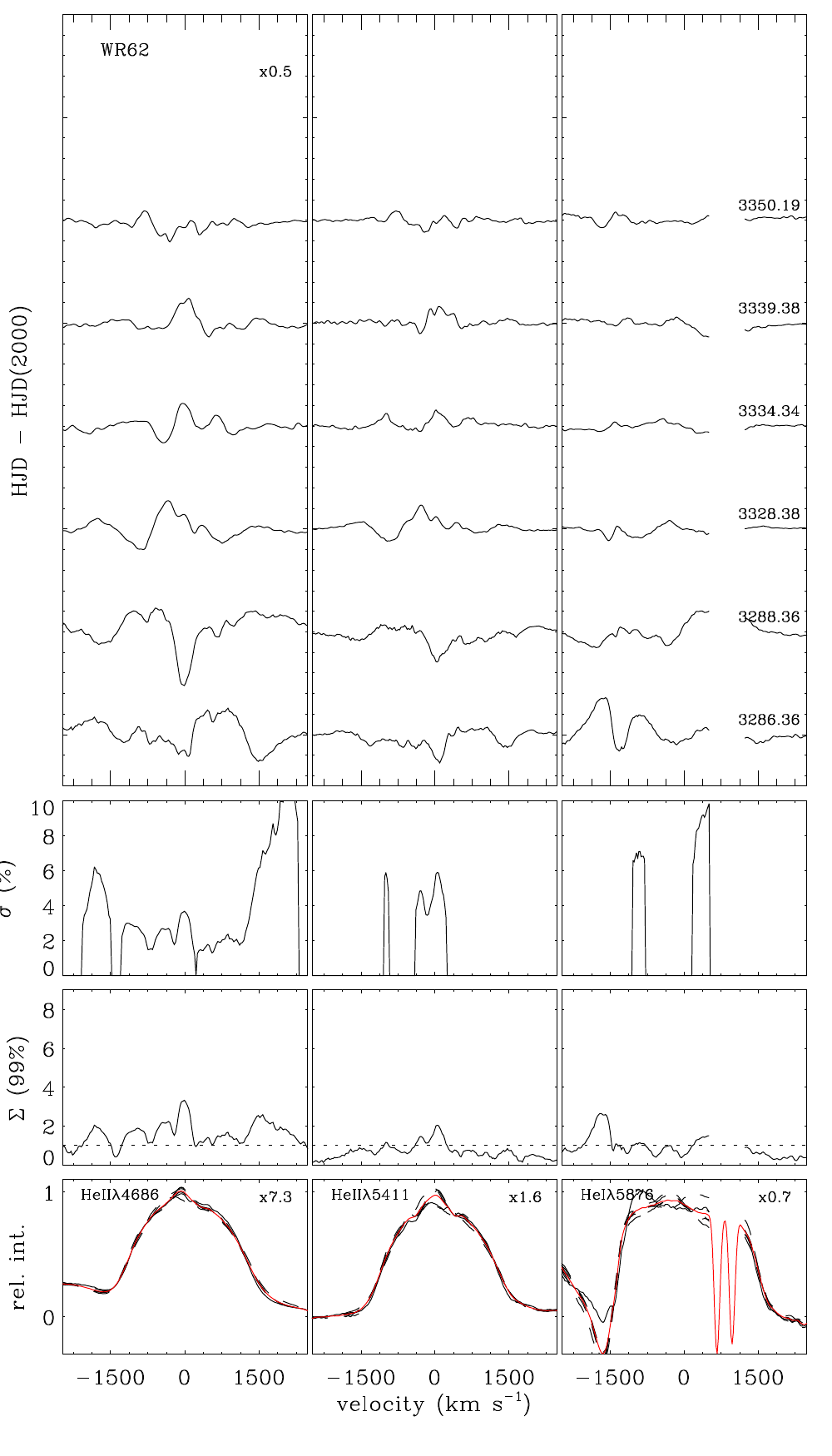}
\end{subfigure}
\begin{subfigure}{.4\textwidth}
%\centering
\includegraphics[width=9cm,height=11cm]{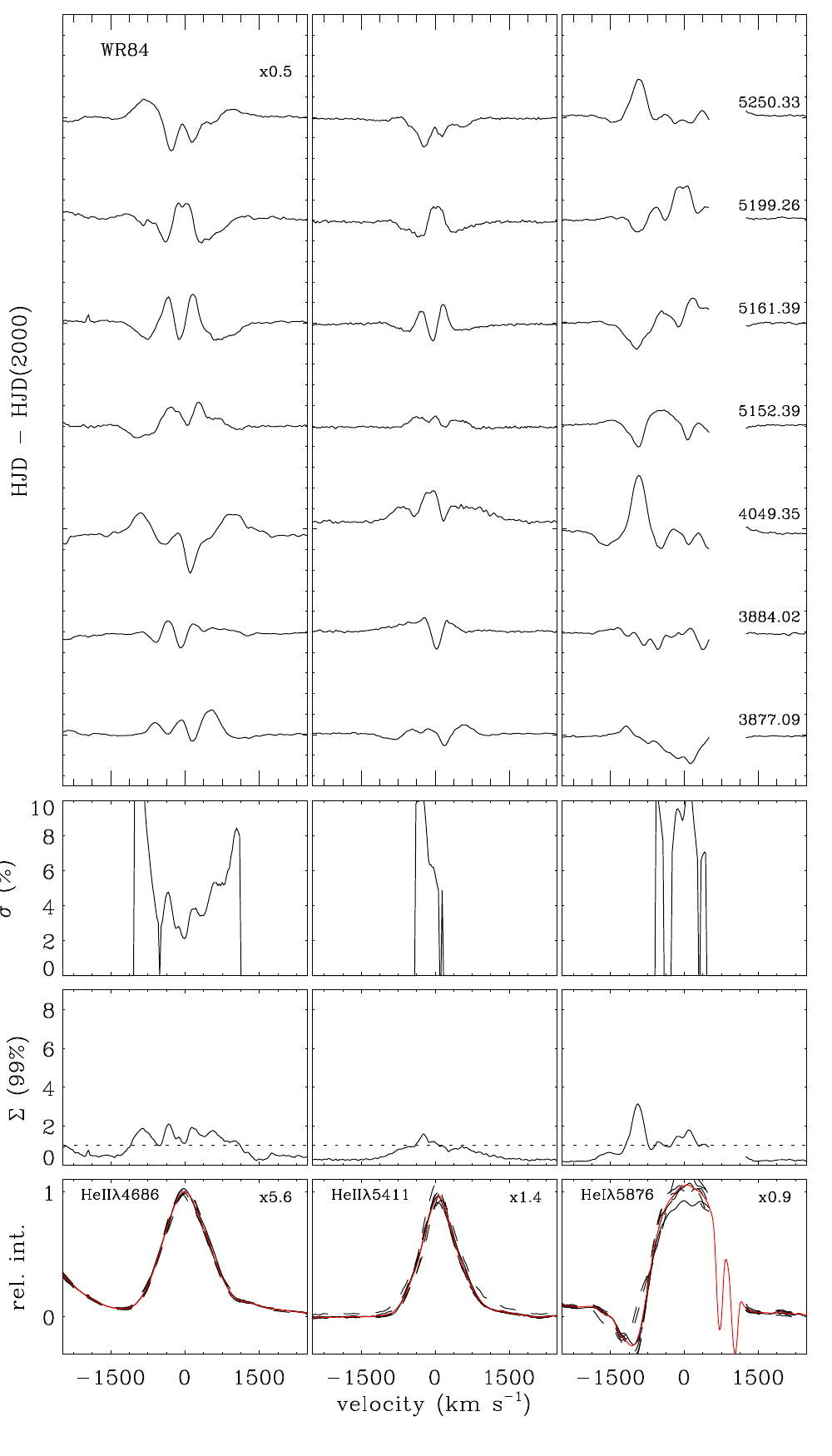}
\end{subfigure}
\begin{subfigure}{.4\textwidth}
%\centering
\hspace{2.1cm}
\includegraphics[width=9cm,height=11cm]{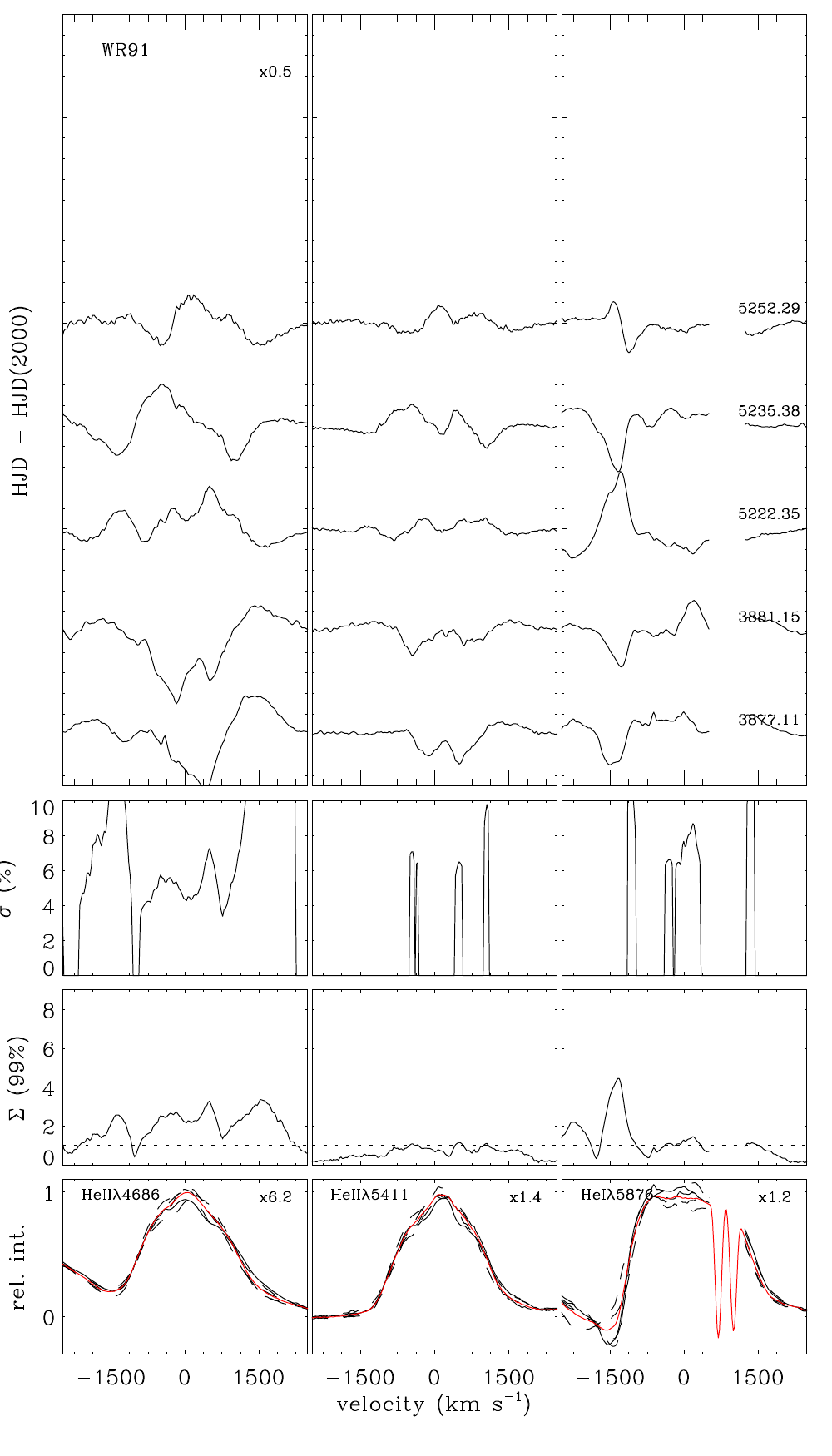}
\end{subfigure}

\caption{ Same as Fig.A.1, but for WR51, WR62, WR84, and WR91.  
}
\end{figure*}

\end{document}